\documentclass[journal]{IEEEtran}
\usepackage{mathrsfs}
\usepackage{amsfonts}
\usepackage{amsthm}
\usepackage{dsfont}
\usepackage{amsopn}
\theoremstyle{remark}
\newtheorem{definition}{Definition}

\newtheorem{theorem}{Theorem}
\newtheorem{lemma}{Lemma}
\newtheorem{corollary}{Corollary}
\newtheorem{remark}{Remark}

\usepackage{algorithm}
\usepackage{color}
\usepackage{cite}
\usepackage{verbatim}

\usepackage{url}
\usepackage{subcaption}

\newcommand{\R}{\ensuremath{\mathbb{R}}}

\DeclareMathOperator{\Id}{\mathbf{I}}

\def\0{\mathbf{0}}
\def\1{\mathds{1}}

\def\x{\mathbf{x}}

\def\y{\mathbf{y}}

\def\u{\mathbf{u}}

\def\w{\mathbf{w}}
\def\p{\mathbf{p}}

\def\S{\mathcal{S}}

\def\N{\mathcal{N}}

\def\V{\mathcal{V}}
\def\M{\mathcal{M}}

\def\D{\mathcal{D}}

\def\Pd{\mathcal{P}}
\def\Ep{\mathbb{E}}
\def\one{{\bf 1}}
\def\V{\mathcal{V}}
\def\G{\mathcal{G}}

\def\E{\mathcal{E}}
\def\O{\mathcal{O}}
\def\Za{\mathbb{Z}}
\def\l{\left}
\def\r{\right}

\ifCLASSINFOpdf
   \usepackage[pdftex]{graphicx}
\else
   \usepackage[dvips]{graphicx}
   \graphicspath{{../eps/}}
   \DeclareGraphicsExtensions{.eps}
\fi

\usepackage[cmex10]{amsmath}
\usepackage{algorithmic}
\begin{document}
%
\title{Fast Path Localization on Graphs via Multiscale Viterbi Decoding}

\author{
\IEEEauthorblockN{Yaoqing Yang\IEEEauthorrefmark{1}\IEEEauthorrefmark{2}, Siheng Chen\IEEEauthorrefmark{1}, Mohammad Ali Maddah-Ali\IEEEauthorrefmark{2}, Pulkit Grover\IEEEauthorrefmark{1}, Soummya Kar,\IEEEauthorrefmark{1} and Jelena Kova\v{c}evi\'{c}}\IEEEauthorrefmark{1}\\
 \IEEEauthorblockA{\IEEEauthorrefmark{1}Carnegie Mellon University}
 \IEEEauthorblockA{\IEEEauthorrefmark{2}Nokia Bell Labs\vspace{-8mm}}

\thanks{A preliminary version of this work was presented in part at the IEEE International Conference on Acoustics, Speech and Signal Processing (ICASSP), 2017 \cite{yang2017fast}. This work is supported in part by the National Science Foundation under grants CCF-1513936, by ECCS-1343324 and CCF-1350314 (NSF CAREER) for Pulkit Grover, and by Systems on Nanoscale Information fabriCs (SONIC), one of the six SRC STARnet Centers, sponsored by MARCO and DARPA.}


}%


%


\maketitle
\begin{abstract}
  We consider a problem of localizing a \emph{path-signal} that evolves over time on a graph. A \emph{path-signal} can be viewed as the trajectory of a moving agent on a graph in several consecutive time points. Combining dynamic programming and graph partitioning, we propose a path-localization algorithm with significantly reduced computational complexity. We analyze the localization error for the proposed approach both in the Hamming distance and the destination's distance between the path estimate and the true path \textcolor{black}{using numerical bounds}. \textcolor{black}{Unlike usual theoretical bounds that only apply to restricted graph models, the obtained numerical bounds apply to all graphs and all non-overlapping graph-partitioning schemes.} In random geometric graphs, we are able to derive a closed-form expression for the localization error bound, and a tradeoff between localization error and the computational complexity. Finally, we compare the proposed technique with the maximum likelihood estimate under the path constraint in terms of computational complexity and localization error, and show significant speedup (100$\times$) with comparable localization error (4$\times$) on a graph from real data. Variants of the proposed technique can be applied to tracking, road congestion monitoring, and brain signal processing.
\end{abstract}

%
\IEEEpeerreviewmaketitle





%

\section{Introduction}
Data with unstructured forms and rich types are being generated from various sources, from social networks to biological networks, from citation graphs to knowledge graphs, from the Internet of Things to urban mobility patterns. These data are often generated with some inherent dependencies that can be represented using graphs and thus inspired the emerging field of graph signal processing~\cite{ShumanNFOV:13, SandryhailaM:14}. In graph signal processing, signals are supported on graphs, instead of on conventional regular well-ordered domains (e.g., signals supported on the time grid or signals supported on other regular grids). This key difference spurred a lot of research that aims to generalize classical techniques to graph signal processing, including sampling~\cite{AnisGO:15, ChenVSK:15, ChenVSK:15c}, recovery~\cite{ChenSMK:14, ChenCRBGK:13, NarangGO:13}, signal representations~\cite{ZhuM:12, ThanouSF:14, ChenVSK:15h}, uncertainty principles~\cite{AgaskarL:13, TsitsveroBL:15}, and graph signal transforms~\cite{NarangSO:10, HammondVG:11,ShumanRV:15,teke2016extending,teke2016extending2}.

\begin{figure}
  \centering
  \includegraphics[scale=0.25]{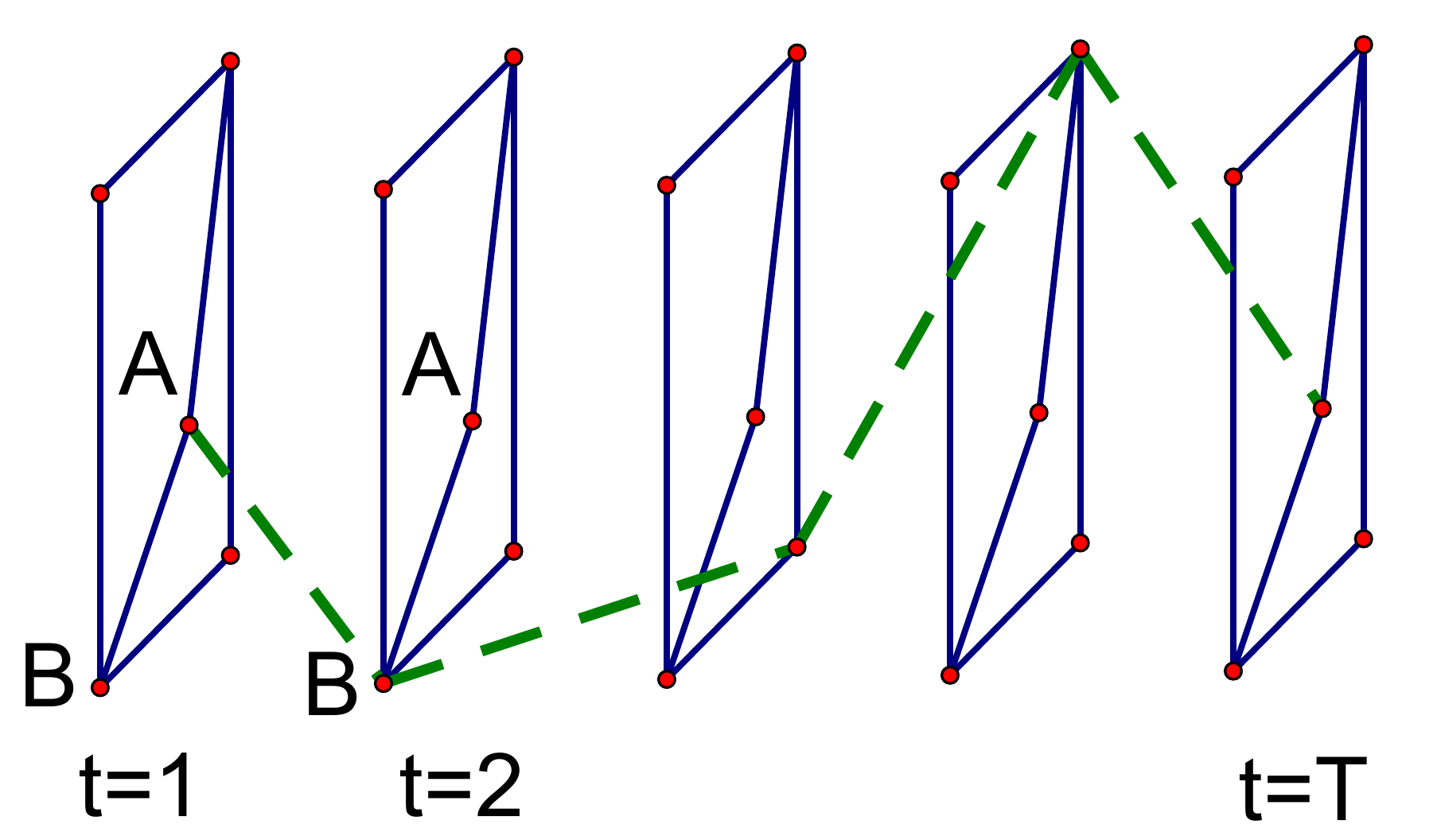}\\
  \caption{A path-signal on a graph with five nodes. The nodes with a non-zero signal value form a (connected) path on the graph (green dashed line). For example, the activated node at time $t=1$ is $v_1=A$ and the activated node at time $t=2$ is $v_2=B$. For the signal to be a path-signal we require $(v_1,v_2)\in\E$. A path-signal can be viewed as an abstraction of the trajectory of a moving agent on a graph.\vspace{-3mm}}\label{fig:graph_series}
\end{figure}
In this paper, we study a special type of \textcolor{black}{dynamic signals} on graphs that we call \emph{path-signals}. A path-signal (see Fig.~\ref{fig:graph_series} for an illustration) is a special type of a graph signal that is supported on a connected trajectory, i.e., the signal is non-zero at only one location at each time point, and the non-zero locations at consecutive time points form a connected path on the graph. A path-signal is an abstraction of a moving agent on a graph, where the non-zero location of the path-signal at a particular time point $t$ can be viewed as the location of the moving agent on the graph at time $t$. Thus, the study of path-signals is deeply related to tracking and surveillance~\cite{oh2005tracking}.

Here, we study the path-signals on large-scale graphs from the perspective of graph partitioning and graph (signal) dimension reduction: due to the increasing size of graphs, many techniques on dimension reduction for graphs or graph signals have been proposed, which include community detection and clustering on graphs~\cite{Tremblay:14, DongFVN:14, ChenO:14,lim2014slashburn} as well as signal coarsening on graphs \cite{lafon2006diffusion,shuman2016multiscale,liu2014graph}. These newly proposed techniques and related ideas have provided great improvements in computation speed and storage cost for algorithms on large-scale graphs, including PageRank \cite{jung2016random}, graph generation \cite{leskovec2010kronecker} and graph semantic summarization \cite{koutra2015summarize}. By studying path-signals, we will be able to explore the connections between signal tracking and graph dimension reduction.

In the path-signal problem, we consider two different subproblems: the ``path-localization'' problem and the ``destination-localization'' problem. The aim of the first problem is to estimate the trajectory of the moving agent, while the aim of the second problem is to estimate only the final position of the moving agent, both from noisy observations of the path-signal. We measure the accuracy of the first problem using the metric of Hamming distance. For the second problem, we measure the path-localization accuracy using the Euclidean distance, assuming the graph is embedded in an Euclidean space (e.g., a geometric graph). First, we propose to use an algorithm based on dynamic programming to estimate the trajectory and destination of the path-signal on the original graph. \textcolor{black}{This algorithm resembles the classical Viterbi decoding method in the context of convolutional decoding \cite{viterbi1967error}}. We also show that this algorithm is the maximum likelihood (MLE) estimate under the path constraint.

The computational complexity of path MLE is high for large graphs, which motivates us to design a fast approximate algorithm. We use graph partitioning techniques to divide the graph into non-overlapping clusters and merge each cluster into a single ``super-node''. Subsequently, we implement dynamic programming on the resulting graph defined by these super-nodes. Since we track the path trajectory and path destination on the graph defined by these clusters, we significantly reduce the number of states in dynamic programming, and hence reduce the computation time. \textcolor{black}{The proposed method can be viewed as implementing the Viterbi decoding method on a condensed graph and hence our approach is referred to as multiscale Viterbi decoding.} Using large-deviation techniques, we provide bounds on the two distance measures (Hamming distance and the destination distance) respectively for the path-localization problem and the destination-localization problem. We show that both bounds can be computed in polynomial time for general graphs and general non-overlapping graph partitioning algorithms.

Then, we focus on an important class of graphs, i.e., random geometric graphs with a simple square tessellation partitioning. The random geometric graph is widely used for sensor networks \cite{bettstetter2002minimum}, and thus is particularly relevant for the study of tracking. In this case, we obtain a closed-form theoretical bound on the localization error. We validate the multiscale Viterbi decoding algorithm on synthesized random geometric graphs by showing both theory and simulation results.

Next, we consider real graph data coming from the autonomous systems in Oregon \cite{leskovec2005graphs} and apply the multiscale Viterbi decoding algorithm with several well-known graph partitioning schemes. The graph partitioning scheme that achieves the best performance is the Slashburn algorithm \cite{lim2014slashburn}, which is based on the idea that real networks have no good cuts (i.e., the vertices of the graph cannot be easily partitioned into two non-overlapping groups such that the number of edges that span the two groups is small) unless a small set of ``hub-nodes'' with high degree is removed. Our algorithm using graph partitioning shows significant speedup and comparable localization performance with respect to the direct method that uses dynamic programming on the original graph without using graph partitioning.

\textcolor{black}{A closely related line of work considers the problem of detecting signals in irregular domains \cite{CastroCD:11,HuCSFLL:13, SharpnackRS:13, SharpnackKS:13, SharpnackKS:13a, ChenYZSK:16}. In particular, \cite{agaskar2013detecting,ting2006near} consider optimal random walk detection on a graph which is closely related to the problem of path localization. However, our problem setting considers path-signals that can be adversarial, in the sense that the proposed algorithm and theoretical bounds can be applied to worst-case path-signals. Moreover, we consider approximate algorithms that have low computational complexity compared to the optimal localizers (such as the one based on MLE that is presented in Section~\ref{sec:dp_general}) but have comparable performance. }

The problem of path localization on graphs is strongly connected to signal tracking. If the graph is viewed as the state space and the Markov transition probability is imposed by the graph topology, the path-localization problem is equivalent to tracking. However, as we mentioned earlier, compared to tracking problems including Kalman filtering \cite{SinopoliSFPJS:04} and particle filtering \cite{GordonN:93}, the proposed method applies to worst-case trajectory and signal. Some other related tracking problems include natural video tracking \cite{lucas1981iterative}, cell tracking \cite{hadjantonakis2004dynamic} and diffusion tensor imaging \cite{le2001diffusion}. Those methods are often applied to regular signals such as time signals or images, but our paper considers signals supported on graphs.

\textcolor{black}{The proposed path-localization problem is related to tracking and trajectory recovery in many different contexts}, such as road congestion monitoring, satellite searching and brain signal processing \cite{betzel2016multi}. A signal path on a road network can be viewed as a slowly moving congested segment on the graph formed by roads and intersections. A signal path in a satellite search can be viewed as the trajectory of \textcolor{black}{a plane debris moved by ocean currents }or the trajectory path of a small refugee lifeboat in the sea, while the observation noise may come from sensing inaccuracy and poor illumination conditions at night. The signal path in a brain imaging problem can be viewed as consecutive firing events of brain signals in the brain network.

\section{System Model and Problem Formulation: Path-Signal and Path Localization}\label{sec:sys_model}

We denote by $\G=(\V,\E)$ an undirected\footnote{The proposed algorithms in this paper naturally apply to directed graphs as well. For the sake of consistency, we only consider undirected graphs in this paper.} graph with $\V$ as the set of nodes and $\E$ the set of edges, where $|\V|=n$ for some $n\in\mathbb{Z}^+$. We use $\x_t\in \R^n,t=1,2, \ldots  T$ to denote a deterministic (but unknown) time-series of (non-random) signals supported on the graph $\G=(\V,\E)$ that evolve over time. The value $\x_t(v)$ denotes the signal value at time $t$ and at node $v$.

\begin{definition}\label{def:path_signal}
\textcolor{black}{(Path-signal) A deterministic but unknown time-series of non-random signals $\x_t\in \R^n,t=1,2, \ldots  T$ is called a path-signal on the graph $\G=(\V,\E)$ if at each time point $t$, there is one node $v_t$ such that $\x_t(v_t)=\mu>0$ and $\x_t(v)=0$ for all other nodes $v\neq v_t$. The collection of the nodes $(v_1,v_2, \ldots v_T)$ forms a \emph{connected path}, i.e., $(v_t,v_{t+1})\in \E$ for all $t=1,2, \ldots T-1$.}
\end{definition}

A path $(v_1,v_2, \ldots v_T)$ can represent the trajectory of a moving agent on the graph $\G=(\V,\E)$ from time $t=1$ to time $t=T$. We call the sequence $\{\x_t\}_{t=1}^T$ the path-signal, since the signal value $\x_t(v)$ on the path $(v_1,v_2, \ldots v_T)$ has a shifted value $\mu>0$.

Let $\p^*=(v^*_1,v^*_2, \ldots v^*_T)$ be a connected path let $\{\x_t\}_{t=1}^T$ be a path signal. Let $\{\y_t\}_{t=1}^T$ be a sequence of noisy observations of a path-signal $\{\x_t\}_{t=1}^T$ where,
\begin{align}
\label{eqn:path_est}
  \y_t =\x_t+\w_t,t=1,2, \ldots T,
\end{align}
where $\w_t\sim \N( \0, \sigma^2 \Id_{n\times n})$ is Gaussian noise. Our goal is to localize this connected path $\p^*=(v^*_1,v^*_2, \ldots v^*_T)$ on the graph $\G=(\V,\E)$ with shifted mean $\mu$ from the noisy observations $\{\y_t\}_{t=1}^T$. We will call $\p^*$ the ``true path''. \textcolor{black}{We will use $\hat{\p}=(\hat{v}_1,\hat{v}_2,\ldots \hat{v}_T)$ to denote the chain estimate.}

\subsection{Two Error Metrics for Path Localization}
We define two different error metrics on the path-localization problem. Using these two metrics, we can measure the inaccuracy of different path-localization algorithms.

\begin{definition}\label{def:Hamming}
(Hamming distance) The Hamming distance between the estimated chain $\hat{\p}=(\hat{v}_1,\hat{v}_2, \ldots \hat{v}_T)$ and the true path $\p^*=(v^*_1,v^*_2, \ldots v^*_T)$ is defined as
\begin{equation}\label{eqn:Hamming}\small
  D_H(\hat{\p},\p^*)=\sum_{t=1}^T  \1(\hat{v}_t\neq v^*_t),
\end{equation}
\end{definition}
where $\1(\cdot)$ denotes the indicator function.
\begin{definition}\label{def:destination}
(Destination distance) The destination distance $D_F(\hat{\p},\p^*)$ between the estimated chain $\hat{\p}=(\hat{v}_1,\hat{v}_2, \ldots \hat{v}_T)$ and the true path $\p^*=(v^*_1,v^*_2, \ldots v^*_T)$ is defined as
\begin{equation}\label{eqn:Dest}\small
D_F(\hat{\p},\p^*)=d(\hat{v}_T,v^*_T),
\end{equation}
where the distance $d(\hat{v}_T,v^*_T)$ is a distance metric between the two nodes $\hat{v}_T$ and $v^*_T$ defined for the graph $\G$ (which can either be the multi-hop distance\footnote{The multi-hop distance between two nodes on an undirected graph is the minimum number of hops required to reach one node from the other through a chain of edges.} on a general graph or the \textcolor{black}{Euclidean distance on a geometric graph \cite{penrose2003random} embedded in an Eclidean space)}.
\end{definition}
The Hamming distance measures the inaccuracy of the chain estimate by counting the overall number of mismatched nodes, while the destination distance directly measures the distance between the final positions of the chain estimate and the true path. The second metric is more useful when the goal is to make an estimate of the current position of the moving agent. We also call the path estimation problem with the destination distance metric the \emph{destination-localization} problem. \textcolor{black}{Here, we only consider localizing each position $v^*_t$ given the entire signal $\{\y_t\}_{t=1}^T$. A generalized problem is that we want to localize each position $v^*_t$ using only the observations up to time $t$, i.e., $\{\y_\tau\}_{\tau=1}^t$.}

\subsection{Constrained Maximum Likelihood Chain Estimators}

Denote by $\V^T$ the $T$-fold Cartesian product of the node set $\V$. Note that $\V^T$ is the set of all possible chains of nodes of length $T$.

\begin{definition}($\S$-constrained MLE)
For an arbitrary set $\S\subset \V^T$, define the $\S$-constrained maximum likelihood estimate (MLE) $\hat{\p}_\S^\text{MLE}$ as the chain in $\S$ that has the maximum likelihood value for the observed signal $\y_t,t=1,\ldots T$. That is
\begin{equation}\small
\p_\S^\text{MLE}=\arg\max_{\hat{\p}\in \S}\Pr\l((\y_1,\dots,\y_T\r)|\p^*=\hat{\p}).
\end{equation}
\end{definition}

First, we show that the $\S$-constrained MLE has the maximum sum signal over all chains in $\S$. For an arbitrary chain of nodes $\hat{\p}=(\hat{v}_1,\hat{v}_2, \ldots \hat{v}_T)\in \S$, we define the sum signal
\begin{equation}\label{eqn:sum_weight}\small
  S(\hat{\p})=\sum_{t=1}^T \y_t(\hat{v}_t).
\end{equation}
Intuitively, a chain with a higher sum signal\footnote{\textcolor{black}{Note that we want to localize the position $v_t$ of the path at each particular time point $t=1,2,\ldots T$, instead of all positions that the path has passed through aggregated in a single snapshot. For the problem of aggregating all locations that the path has passed through in a single snapshot, we may simply compute the sum of all signals $\sum_{t=1}^T \y_t$ in all time points and select the nodes $v$ with a higher sum signal.}} is more likely to be the true path $\p^*$. In fact, it coincides with the $\S$-constrained MLE, as shown below:
\[\small\begin{split}
\Pr\l((\y_1,\dots,\y_T\r)|\p^*)=\prod_{t=1}^T C\exp \l(-\frac{1}{2\sigma^2}\l\|\y_t-\x_t\r\|_2^2\r)\\
=C^T\exp\l(-\frac{1}{2\sigma^2}\sum_{t=1}^T\l\|\y_t-\x_t\r\|_2^2\r),
\end{split}\]
where recall that $\p^*=(v^*_1,v^*_2,\ldots v^*_T)$ denotes the true path and $C$ is a constant. Then,
\begin{equation}\label{eqn:path_sum_est}
\small\begin{split}
&\arg\max_{\hat{\p}\in\S}\Pr\l((\y_1,\dots,\y_T\r)|\p^*=\hat{\p}\in\S)=\arg\min_{\hat{\p}\in\S}\sum_{t=1}^T\l\|\y_t-\x_t\r\|_2^2\\
=&\arg\max_{\hat{\p}\in\S}\sum_{t=1}^T\y_t\cdot\x_t=\arg\max_{\hat{\p}\in\S}\sum_{t=1}^T \mu\y_t(v_t)\\
=&\arg\max_{\hat{\p}=(\hat{v}_1,\hat{v}_2,\ldots \hat{v}_T)\in\S}\sum_{t=1}^T\y_t(\hat{v}_t).
\end{split}
\end{equation}

Then, we consider two extreme cases of $\S$-constrained MLE. In the first case, we set $\S=\V^T$. In this case, the $\S$-constrained MLE is a naive estimator that completely ignores the path constraint. In fact, since the signals on each time point is independent of each other, the $\S$-constrained MLE is $\hat{\p}_{\V^T}^\text{MLE}=(\hat{v}_1,\hat{v}_2,\ldots \hat{v}_T)$, where
\begin{equation}\label{eqn:naive_max}\small
\hat{v}_t=\arg\max_{\hat{v}\in\V}{\;\;} \y_t(\hat{v}),t=1,2,\dots,T.
\end{equation}
Although this estimator is extremely simple, it does not perform well respect to the proposed distance metrics in Definition \ref{def:Hamming} and \ref{def:destination} due to ignoring the path constraint (we will show this later in Section~\ref{sec:hub_graph}).

The second $\S$-constrained MLE is the maximum likelihood estimator under the constraint that the estimator $\hat{\p}=(\hat{v}_1,\hat{v}_2,\ldots \hat{v}_T)$ has to be a path. We describe this particular $\S$-constrained MLE in the following section.

\subsection{Path-constrained MLE using Viterbi Decoding}\label{sec:dp_general}
From \eqref{eqn:path_sum_est}, if we impose the constraint that $\hat{\p}$ must be a connected path, the MLE estimate of $\p^*$ is the connected path with the maximum sum signal. In Algorithm~\ref{alg:dp}, we describe a dynamic programming algorithm to compute a connected path with the maximum sum signal. This algorithm is also known as the Viterbi decoding algorithm in the context of convolutional decoding \cite{viterbi1967error}. The basic idea in Algorithm~\ref{alg:dp} is to record the path with the largest sum signal of length $t$ that ends at node $v$ for all nodes in the graph $\G$ and all time points $t=1,2, \ldots T$. Although the possible number of paths is exponential in $t$, Algorithm~\ref{alg:dp} has computational complexity $\mathcal{O}(nT)$, because only the optimal path, instead of all paths, that ends at a node $v$ has to be recorded at each time $t$.
\begin{algorithm}\caption{Dynamic Programming for Path-Signal Localization}\label{alg:dp}
\textbf{INPUT}: A graph $\G=(\V,\E)$ and graph signal observations $\y_t,t=1,2, \ldots T$.\\
\textbf{OUTPUT}: A chain of nodes $\hat{\p}=(\hat{v}_1,\hat{v}_2, \ldots \hat{v}_T)$.\\
\textbf{INITIALIZE} \\
Use $s_{v,t}$ to denote the sum signal until time $t$ at node $v$. Use $\p_{v,t}$ to denote the path with the largest sum signal of length $t$ that ends at node $v$. Initialize $\p_{v,1}=v$ for all $v\in\V$.\\
\textbf{FOR} $t$=$2:T$
\begin{itemize}
  \item For all nodes $v$ in $\V$, let $ u_m= \arg\max_{u\in\mathcal{N}(v)}S(\mathbf{p}_{u, t-1})$, where $\mathcal{N}(v)$ denotes the neighborhood of $v$. Therefore, the path $\p_{u_m,t-1}$ has the largest sum signal $S(\p_{u,t-1})$ for all nodes $u$ in the neighborhood $\N(v)$;
  \item Update $\p_{v,t}=(\p_{u_m,t-1},v)$ for all $v\in\V$.
\end{itemize}
\textbf{END}\\
Denote by $\p_{v^*,T}$ the path with the largest sum signal in all paths of length $T$. Output $\hat{\p}=\p_{v^*,T}$.
\end{algorithm}

The $\S$-constrained MLE with $\S$ being all connected paths in $\V^T$ has much better empirical performance than the naive maximization $\hat{\p}_{\V^T}^\text{MLE}$ in \eqref{eqn:naive_max} (we will show this later in Section~\ref{sec:hub_graph}). However, the computational complexity $\mathcal{O}(nT)$ is high for a large graph and a large overall time $T$. This motivates us to design some approximate algorithms that have low computational complexity (see Section~\ref{sec:dp_app}).

\begin{remark}\label{rmk:not_path}
Note that in the path-localization problem, we do not require the estimated chain of nodes $\hat{\p}=(\hat{v}_1,\hat{v}_2,\ldots \hat{v}_T)$ to be connected, i.e., $(\hat{v}_t,\hat{v}_{t+1})\in\E,t=1,2,\ldots T-1$. As long as we have an estimated chain of nodes with small Hamming distance or destination distance as defined in Definition \ref{def:Hamming} and \ref{def:destination}, the estimate is good. The naive $\V^T$-constrained MLE in \eqref{eqn:naive_max} is bad, not because it is disconnected, but because it totally ignores the path constraint and results in large Hamming distance. The path-constrained MLE is good in empirical performance but it has high computational complexity. To resolve this issue, we propose an approximate path estimate in Section \ref{sec:dp_app}, which is not necessarily connected in the original graph $\G$, but connected in a graph formed by subgraphs (we will present the details in Section\ref{sec:dp_general}). Moreover, it has lower computational complexity (100+ times speed-up) than the path-constrained MLE, and comparable localization error (4$\times$ in Hamming distance). Therefore, the approximate path estimate can be viewed as a relaxed version of path-constrained MLE, with lower computational complexity.
\end{remark}

\section{Multiscale Viterbi Decoding for Fast Path-Signal Localization}\label{sec:dp_app}
In this section, we design an algorithm that combines path localization with graph partitioning. As we have mentioned in Section~\ref{sec:sys_model}, we want to use an estimation technique that has very low computational complexity. Our main idea is to first partition the original graph into clusters~\cite{fortunato2010community} and localize an approximate path on a new graph formed by the clusters\footnote{The graph partitioning that we consider only means the partitioning of the node set $\V$ into non-overlapping subsets. It does not necessarily correspond to the usual definition of graph partitioning, i.e., the number of intra-cluster edges is greater than number of inter-cluster edges and the number of nodes in different clusters should be similar.}. \textcolor{black}{Then, we do a refined search inside the approximate path to search for a chain of nodes that well approximates the true path in the original graph in terms of the Hamming distance and the destination distance.}

\begin{figure}
  \centering
  \includegraphics[scale=0.17]{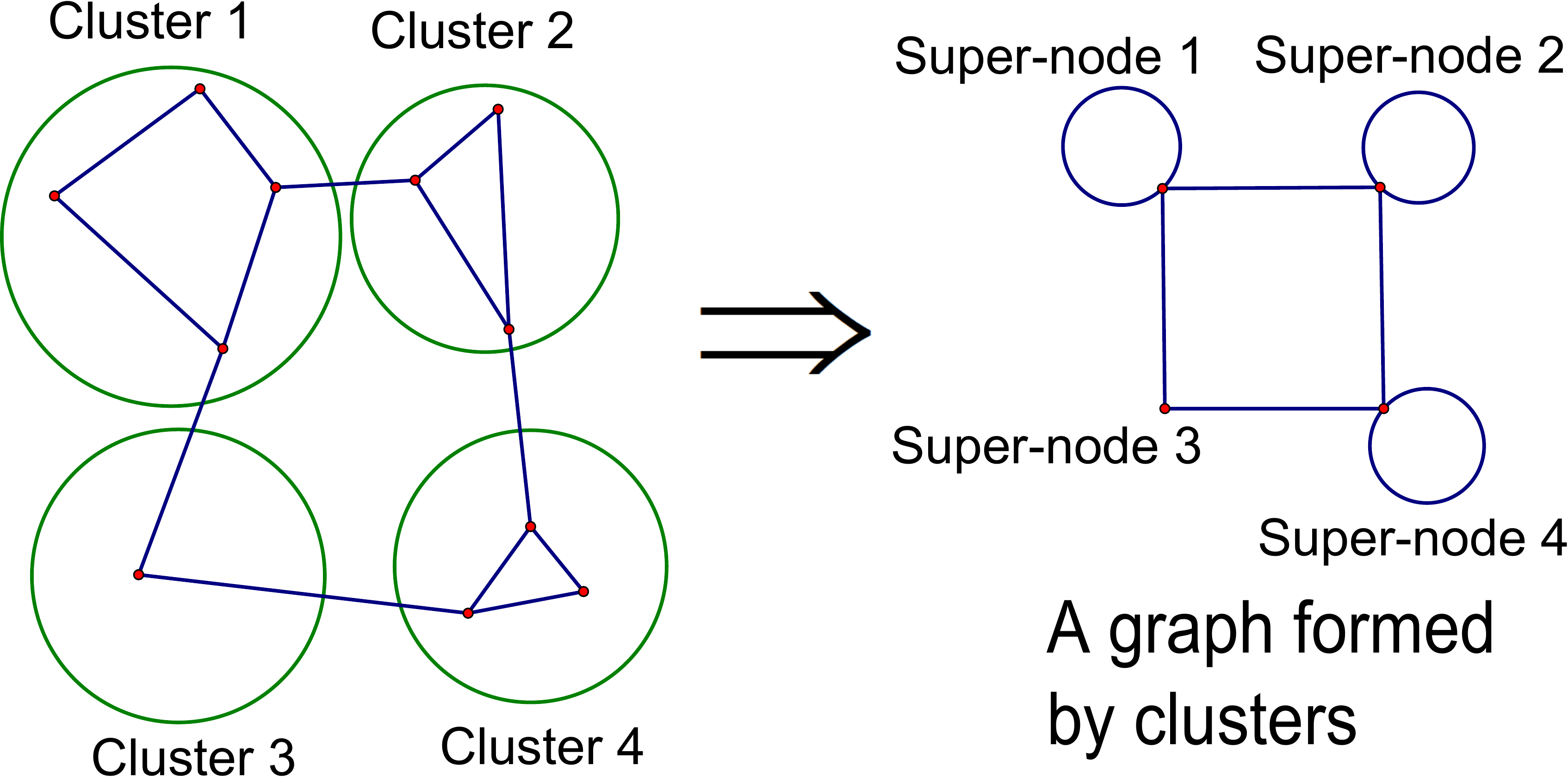}\\
  \caption{An illustration of the graph formed by clusters. In the original graph $\G$, we partition the nodes into non-overlapping clusters. Then, we shrink each cluster to one ``super-node'' and connect two super-nodes if there exist two connected nodes in the corresponding two clusters. The ``super-graph'' may have self-loops but our algorithms still apply.\vspace{-3mm}}\label{fig:graph_community}
\end{figure}

Suppose we partition the nodes of the graph $\G=(\V,\E)$ into $m$ non-overlapping clusters
\begin{equation}\label{eqn:community_graph}\small
\V=\bigcup_{i=1}^m \V_i.
\end{equation}
Then, we shrink each cluster into a ``super-node'' and construct a new graph denoted by $\G^\text{new}=(\V^\text{new},\E^\text{new})$ formed by these super-nodes as follows. The node set $\V^\text{new}$ with cardinality $|\V^\text{new}|=m$ is the set of ``super-nodes''. Two nodes $\V_i$ and $\V_j$ are connected if there exists two nodes $v_i\in \V_i$ and $v_j\in \V_j$ such that $(v_i,v_j)\in\E$ in the original graph $\G$. We call the graph $\G^\text{new}$ the super-graph (see Fig.~\ref{fig:graph_community}).

Consider the observation model described in~\eqref{eqn:path_est} on a general graph $\G=(\V,\E)$ such that $\x_t(v_t)=\mu$ and $\x_t(v)=0$ for $v\neq v_t$, and $(v_1,v_2, \ldots v_T)$ is a connected path in $\G$. Suppose we use a graph-partitioning algorithm and obtain the super-graph $\G^\text{new}=(\V^\text{new},\E^\text{new})$. We use a coarsened \cite{lafon2006diffusion,shuman2016multiscale,liu2014graph} version $\u_t$ of the original observation $\y_t$ as the graph signal on the super-graph. The coarsened graph signal is defined as
\begin{equation}\label{eqn:signal_coarsening}\small
  \u_t(\V_i)=\max_{v\in \V_i}\y_t(v), i=1,2, \ldots m.
\end{equation}
We will briefly discuss the reason that we choose the $\max$ statistic instead of other statistics such as average in Remark~\ref{rmk:max_statistic}. Then, using the coarsened signal, we execute the same dynamic programming algorithm as Algorithm~\ref{alg:dp} to obtain an estimate of the trajectory of the path-signal on the super-graph (see Algorithm~\ref{alg:dp_app}).
\begin{algorithm}\caption{Coarsened Dynamic Programming for Path-Signal Localization}\label{alg:dp_app}
\textbf{INPUT}: A coarsened graph $\G^\text{new}=(\V^\text{new},\E^\text{new})$ and coarsened graph signal observations $\u_t,t=1,2, \ldots T$.\\
\textbf{OUTPUT}: A chain of nodes $\hat{\Pd}=(\hat{\V}_1,\hat{\V}_2, \ldots \hat{\V}_T)$ on the super-graph.\\
Call Algorithm~\ref{alg:dp} with inputs $\G^\text{new}=(\V^\text{new},\E^\text{new})$ and $\u_t,t=1,2, \ldots T$.
\end{algorithm}
Note that after graph partitioning, the sum signal maximization does not equal to the MLE because the signal $\u_t(\V_i)$'s are not Gaussian. However, we will show an upper bound on the expectation of the Hamming and destination distance between the true path and the path estimate on the super-graph $\G^\text{new}$ (see Section \ref{sec:analysis}). Therefore, Algorithm~\ref{alg:dp_app} is an approximate path localization algorithm that aims to reduce computational complexity of MLE.
\begin{remark}\label{rmk:max_statistic}
Choosing the max statistic instead of the mean statistic is reminiscent of the generalized likelihood ratio test (GLRT) \cite{zeitouni1992generalized} for composite hypothesis testing. Although our idea of choosing the max statistic is influenced by GLRT, our localization algorithm is not GLRT. Another reason that we choose the $\max$ statistic instead of the mean statistic for approximate path localization is that the $\max$ statistic has a better localization error than the mean statistic. Consider a cluster $\V$ with one activated node $v_0$. The mean statistic is $u_{\text{mean}}=\frac{1}{|\V|}\sum_{v\in \V}\y_t(v)$ and the $\max$ statistic is $u_{\text{max}}=\max_{v\in \V}\y_t(v)$. When the number of nodes $|\V|\to \infty$, $u_{\text{mean}}\to 0$ almost surely, while $u_{\text{max}}=\max_{v\in \V}\y_t(v)$ still provides some information about the activated node. For a finite $|\V|$, as long as the noise variance is small, the $\max$ statistic can cancel the noise sufficiently well and give a coarsened signal that equals $\y_t(v_0)$, the signal at the activated node $v_0$. However, if we choose the mean statistic $u_{\text{mean}}=\frac{1}{|\V|}\sum_{v\in \V}\y_t(v)$, $\y_t(v_0)$ is always averaged by noise, which makes the performance of the multiscale Viterbi Decoding algorithm degrade. The analysis of the $\max$ statistic is established later formally using large-deviation bounds in Section \ref{sec:analysis}. Note that our algorithm and the bounds in Section \ref{sec:analysis} apply to worst-case path-signals and graph partitioning, which is different from the Bayesian settings in \cite{agaskar2013detecting,ting2006near}.
\end{remark}

\textcolor{black}{After executing Algorithm \ref{alg:dp_app}, we obtain an approximate path estimate $\hat{\Pd}=(\hat{\V}_1,\hat{\V}_2, \ldots \hat{\V}_T)$ in the super-graph. In some tracking and surveillance applications, an approximate path estimate is good enough for subsequent actions. However, in most cases we want to obtain an estimate not only in the super-graph, but also in the original graph,} especially when the clusters are large and the exact positions of the agent are required. In this case, we do a refined search in the original graph in the path $\hat{\Pd}=(\hat{\V}_1,\hat{\V}_2,\ldots\hat{\V}_T)$ that we obtained in the super-graph. Therefore, we propose the final multiscale Viterbi decoding method in Algorithm \ref{alg:dp_multi_scale}.

In Supplementary Section~\ref{sec:multi_path}, we extend the Multiscale Viderbi decoding algorithm to the situation when we have more than one path.

\begin{algorithm}\caption{Multiscale Viterbi Decoding for Path Localization}\label{alg:dp_multi_scale}
\textbf{INPUT}: A graph $\G=(\V,\E)$ and graph signal observations $\y_t,t=1,2, \ldots T$.\\
\textbf{OUTPUT}: A chain of nodes $\hat{\p}=(\hat{v}_1,\hat{v}_2, \ldots \hat{v}_T)$.\\
\textbf{INITIALIZE}\\
Call Algorithm~\ref{alg:dp_app} with inputs $\G=(\V,\E)$ and $\y_t,t=1,2, \ldots T$ to obtain a coarse path estimate $\hat{\Pd}=(\hat{\V}_1,\hat{\V}_2,\ldots\hat{\V}_T)$.

Choose $\hat{v}_t$ as the node that has the largest signal in $\hat{\V}_t$, i.e., $\hat{v}_t=\arg\max_{v\in\hat{\V}_t}\y_t(v)$. Let $\hat{\p}=(\hat{v}_1,\hat{v}_2, \ldots \hat{v}_T)$ be the final chain estimate in the original graph.
\end{algorithm}

\begin{remark}\label{rmk:not_path_1}
As we can see from Algorithm \ref{alg:dp_multi_scale}, although the path estimate $\hat{\Pd}=(\hat{\V}_1,\hat{\V}_2,\ldots\hat{\V}_T)$ is a connected path in the super-graph, the chain estimate $\hat{\p}$ is not necessarily a connected path in the original graph because we choose the node $\hat{v}_t$ with the largest signal in $\hat{\V}_t$ without a path constraint. However, as we mentioned in Remark \ref{rmk:not_path}, as long as the distance metric between the estimated chain $\hat{\p}$ and the true path is small, it does not matter whether $\hat{\p}$ is connected or not. Therefore, $\hat{\p}$ can be viewed as an estimate when the path constraint is relaxed.

\textcolor{black}{Define $\Pd^*=(\V^*_1,\V^*_2, \ldots \V^*_T)$ as the true path in the super-graph, i.e., the true path in the original graph $\p^*=(v^*_1,v^*_2,\ldots v^*_T)$ satisfies $v^*_t\in\V^*_t,\forall t$. Note that the Hamming distance between the approximate path estimate $\hat{\Pd}=(\hat{\V}_1,\hat{\V}_2,\ldots\hat{\V}_T)$ (output from Algorithm~\ref{alg:dp_app}) and the true path $\Pd^*$ in the super-graph is always a lower bound on the Hamming distance between the chain estimate $\hat{\p}=(\hat{v}_1,\hat{v}_2,\ldots\hat{v}_T)$ (output from Algorithm~\ref{alg:dp_multi_scale}) and the true path $\p^*$ in the original graph, i.e.,
\begin{equation}\small
D_H(\hat{\Pd},\Pd^*)\le D_H(\hat{\p},\p^*).
\end{equation}
This is because $\hat{\p}$ is constrained inside the approximate path $\hat{\Pd}$, and hence $\hat{\V}_t\neq \V^*_t$ implies $\hat{v}_t\neq v_t$. However, in Section \ref{sec:hub_graph}, we use a simulation result on a real graph (see Fig. \ref{fig:AS} for details) to show that $D_H(\hat{\p},\p^*)$ is only slightly larger than $D_H(\hat{\Pd},\Pd^*)$. This means that even if we impose the connected path constraint when doing the refined search and use other techniques such as a second round of dynamic programming in the subgraph induced from the path estimate obtained from Algorithm \ref{alg:dp_app} to obtain a connected path in the original graph, we cannot gain too much compared to simply choosing the node with the maximum signal in each cluster as we do in Algorithm~\ref{alg:dp_multi_scale}.}

In the next section, we show a numeric way to compute an upper bound on both the Hamming distance and destination distance in Algorithm \ref{alg:dp_app} and Algorithm \ref{alg:dp_multi_scale} in polynomial time (more specifically, linear in the number of clusters $m$ and at most quadratic in the total number of time points $T$).
\end{remark}

\subsection{A Numeric Method for Computing an Upper Bound on the Localization Error}\label{sec:analysis}

First, we introduce some notation for the analysis of the localization error of the multiscale Viterbi decoding algorithm. We use $W^\text{off}$ to denote a Gaussian random variable $\N( \0, \sigma^2)$ and use $U^\text{on}$ to denote a Gaussian random variable $\N( \mu, \sigma^2)$. We use the superscripts ``on'' and ``off'' because the signal on the path has an elevated mean value $\mu$ while the signal off the path has mean value 0. We use $W_s^\text{off}$ to denote a random variable that is the maximum of $s$ i.i.d. Gaussian random variables with the same distribution as $W^\text{off}\sim \N( \0, \sigma^2)$. We use $U_s^\text{on}$ to denote a random variable that is the maximum of one Gaussian random variable with the same distribution as $U^\text{on}\sim \N( \mu, \sigma^2)$ and $s-1$ Gaussian random variables $W^\text{off}\sim \N( \0, \sigma^2)$.

Recall that $\Pd^*=(\V^*_1,\V^*_2, \ldots \V^*_T)$ is the true path in the super-graph and $\hat{\Pd}=(\hat{\V}_1,\hat{\V}_2, \ldots \hat{\V}_T)$ is the path estimate in the super-graph. Note that at some positions, the two paths may overlap, i.e., $\hat{\V}_t=\V^*_t$ for some $t$. Denote by $S(\Pd^*)$ the sum signal of the true path $\Pd^*$ in the super-graph, and denote by $S(\hat{\Pd})$ the sum signal on the path estimate $\hat{\Pd}$ in the super-graph. Then, from \eqref{eqn:signal_coarsening},
\begin{align}
    &\text{Sum signal on the true path $\Pd^*$: } S(\Pd^*)=\sum_{t=1}^T\u_t(\V^*_t),\\
    &\text{Sum signal on the path estimate $\hat{\Pd}$: } S(\hat{\Pd})=\sum_{t=1}^T\u_t(\hat{\V}_t),
\end{align}
where on the true path, $\u_t(\V^*_t)\overset{\D}{=}U_{|\V^*_t|}^\text{on}$ and on the path estimate, if $\hat{\V}_t\neq \V^*_t$, $\u_t(\hat{\V}_t)\overset{\D}{=}W_{|\hat{\V}_t|}^\text{off}$ ($\overset{\D}{=}$ means equal in distribution). Here recall that $U_s^\text{on}$ denotes the maximum of one Gaussian random variable with mean $\mu$ and $s-1$ Gaussian random variables with mean $0$, $W_s^\text{off}$ denotes the maximum of $s$ Gaussian random variables with mean $0$, and $|\V^*_t|$ and $|\hat{\V}_t|$ denote respectively the number of nodes in the cluster that the true path passes through at time $t$, and the path estimate passes through at time $t$. In the dynamic programming Algorithm~\ref{alg:dp_app} in the super-graph, we will select the path with the maximum sum signal. Therefore, we will choose the path estimate $\hat{\Pd}$ with sum signal $S(\hat{\Pd})$ instead of the true path $\Pd^*$ with sum signal $S(\Pd^*)$, only if $S(\hat{\Pd})\ge S(\Pd^*)$. This event happens with exponentially low probability because the signal on the true path has a shifted mean value  $\mu> 0$, while the signal on $\hat{\Pd}$ has mean value $0$ when the two paths do not overlap.
\begin{lemma}\label{lmm:pro_upb}
The probability that the sum signal $S(\hat{\Pd})$ of the path estimate $\hat{\Pd}=(\hat{\V}_1,\hat{\V}_2, \ldots \hat{\V}_T)$ is greater or equal to the sum signal $S(\Pd^*)$ on the true path $\Pd^*=(\V^*_1,\V^*_2, \ldots \V^*_T)$ can be upper bounded by
\begin{equation}\label{eqn:pro_upb}\small
  \Pr(S(\hat{\Pd})\ge S(\Pd^*))\le \prod_{t\in \Delta}\theta\l(\frac{\mu}{2\sigma^2},|\hat{\V}_t|\r),
\end{equation}
where $\Delta\subset\{1,2, \ldots T\}$ is the set of time indices at which $\hat{\V}_t\neq \V^*_t$, and the function $\theta(\cdot,\cdot)$ is defined as
\begin{equation}\label{eqn:funcf}
\begin{split}
&\theta(s,l):=\min_{\eta\in[0,1]}l\eta^{l-1}e^{s^2\sigma^2-\mu s} Q(Q^{-1}(\eta/\sigma)-s\sigma^2)\\
&+\sqrt{ \frac{l^2}{2l-1} (1-\eta^{2l-1})}e^{\frac{3}{2}s^2\sigma^2-\mu s}\sqrt{ Q(Q^{-1}(\eta/\sigma)-2s\sigma^2)},
\end{split}
\end{equation}
where $s\in\mathbb{R}$ and $l\in\mathbb{Z}^+$.
\end{lemma}
\begin{IEEEproof}
\textcolor{black}{See Supplementary Section~\ref{app:pro_upb} for details of the proof. The basic idea of the proof is to use large-deviation techniques to bound the event $S(\hat{\Pd})\ge S(\Pd^*)$, which is equivalently to
\begin{equation}\small
\sum_{t=1}^T\u_t(\V^*_t)\le\sum_{t=1}^T\u_t(\hat{\V}_t),
\end{equation}
where $\u_t(\V^*_t)$ is the coarsened signal on the true approximate path $\Pd^*$ at time $t$ and $\u_t(\hat{\V}_t)$ is the coarsened signal on $\hat{\Pd}$ at time $t$. Both $\u_t(\V^*_t)$ and $\u_t(\hat{\V}_t)$ are maximum of certain Gaussian random variables, but $\u_t(\V^*_t)$ has one Gaussian random variable with an elevated mean value $\mu$, and hence is more likely to be larger than $\u_t(\hat{\V}_t)$. More specifically, using the large-deviation bound and some derivations, we obtain
\begin{equation}\label{eqn:middle_step}\small
\begin{split}
\Pr(S(\hat{\Pd})\ge S(\Pd^*))\le \min_{s>0}\prod_{t\in \Delta}\Ep\l[e^{s\u_t(\hat{\V}_t)}\r]\Ep\l[e^{-s\u_t(\V^*_t)}\r].
\end{split}
\end{equation}
The bound in \eqref{eqn:pro_upb} is obtained by directly upper-bounding the right hand side of \eqref{eqn:middle_step}.
}
\end{IEEEproof}
Using the conclusion of Lemma~\ref{lmm:pro_upb}, we immediately obtain the following results respectively regarding the localization error metrics in Definition~\ref{def:Hamming} and Definition~\ref{def:destination}. The proofs are omitted because they are direct applications of the following union bound
\begin{equation}\label{eqn:union_bound}\small
  \Ep\l[D(\Pd^*,\hat{\Pd})\r]=\sum_{\forall \hat{\Pd}}\Pr(\hat{\Pd}\text{ is the chosen estimate})D(\Pd^*,\hat{\Pd}),
\end{equation}
where $\sum_{\forall \hat{\Pd}}$ is summation over $\text{all possible paths $\hat{\Pd}=(\hat{\V}_1,\hat{\V}_2, \ldots \hat{\V}_T)$}$ in the super-graph, $D(\cdot,\cdot)$ can be either the Hamming distance $D_H(\cdot,\cdot)$ or the destination distance $D_F(\cdot,\cdot)$.
\begin{theorem}\label{thm:Hamming}
(Hamming distance in Algorithm~\ref{alg:dp_app}) The expectation of the Hamming distance between the path estimate and the true path measured on the super-graph is upper bounded by
\begin{equation}\label{eqn:dis_Hamming}\small
\begin{split}
&\Ep\l[D_H(\Pd^*,\hat{\Pd})\r]\le \sum_{\forall \hat{\Pd}}\l[|\Delta(\hat{\Pd})|\prod_{t\in \Delta(\hat{\Pd})}\theta\l(\frac{\mu}{2\sigma^2},|\hat{\V}_t|\r)\r]\\
&\le \min_{\delta\in[0,1]}\delta T +\sum_{\forall \hat{\Pd} \text{ s.t. } |\Delta(\hat{\Pd})|> \delta T}\l[|\Delta(\hat{\Pd})|\prod_{t\in \Delta(\hat{\Pd})}\theta\l(\frac{\mu}{2\sigma^2},|\hat{\V}_t|\r)\r],
\end{split}
\end{equation}
where $\Delta(\hat{\Pd})$ denotes the set of time indices at which the path estimate $\hat{\V}_t$ is wrong, i.e.,
\begin{equation}\label{eqn:def_sp}
\Delta(\hat{\Pd})=\{t\in \{1,2, \ldots T\}:\hat{\V}_t\neq\V^*_t \},
\end{equation}
and $\theta(\cdot,\cdot)$ is defined in \eqref{eqn:funcf}.
\end{theorem}
\begin{theorem}\label{thm:destination}
(Destination distance in Algorithm~\ref{alg:dp_app}) The expectation of the destination distance between the path estimate and the true path measured on the super-graph is upper bounded by
\begin{equation}\label{eqn:dis_destination}\small
\Ep\l[D_F(\Pd^*,\hat{\Pd})\r]\le \sum_{\forall \hat{\Pd}}\l[d(\V^*_T,\hat{\V}_T)\prod_{t\in \Delta(\hat{\Pd})}\theta\l(\frac{\mu}{2\sigma^2},|\hat{\V}_t|\r)\r],
\end{equation}
where $\Delta(\hat{\Pd})\subset\{1,2, \ldots T\}$ is the set of time indices $t$ at which $\hat{\V}_t\neq\V^*_t$, $\theta(\cdot,\cdot)$ is defined in \eqref{eqn:funcf},
and $d(\V^*_T,\hat{\V}_T)$ is the distance metric between the two super-nodes $\V^*_T$ and $\hat{\V}_T$ in the super-graph.
\end{theorem}

The proofs for these two bounds are omitted because they follow directly from \eqref{eqn:union_bound}. The second inequality in \eqref{eqn:dis_Hamming} is obtained by only counting the paths that satisfy $|\Delta(\hat{\Pd})|> \delta T$.

The two bounds in Theorem~\ref{thm:Hamming} and Theorem~\ref{thm:destination} are stated for the distance metric stated in the super-graph. However, for Algorithm \ref{alg:dp_multi_scale}, we have to compute the distance metric in the original graph. Likewise, we will select a path with the maximum sum signal in Algorithm \ref{alg:dp_multi_scale}, so we can upper-bound the probability of choosing a particular path using an event that happens with small probability.

\begin{definition}\label{def:projected_path}
We will call $\Pd=(\V_1,\V_2,\ldots \V_T)$ in the super-graph $\G^\text{new}=(\V^\text{new},\E^\text{new})$ the \emph{projected path} of a path $\p=(v_1,v_2,\ldots v_T)$ in the original graph $\G=(\V,\E)$ if $v_t\in\V_t$ for $t=1,2,\ldots T$.
\end{definition}

For the true path $\Pd^*=(\V^*_1,\V^*_2,\ldots \V^*_T)$, define the ``first-$k$'' sum as
\begin{equation}\small
f(k)=\max_{\{t_1,t_2,\ldots t_k\}\subset [T]}\sum_{i=1}^k\theta\l(\frac{\mu}{2\sigma^2},|\V^*_t|\r),
\end{equation}
where $\theta(\cdot,\cdot)$ is defined in \eqref{eqn:funcf}.

\begin{theorem}\label{thm:Hamming_multiscale}
\textcolor{black}{(Hamming distance in Algorithm~\ref{alg:dp_multi_scale}) Let $\hat{\p}$ be the estimated chain of nodes using Algorithm~\ref{alg:dp_multi_scale} and let $\p^*$ be the true path. Let $\Pd^*$ be the projected path of $\p^*$. The expectation of the Hamming distance between $\hat{\p}$ and the true path $\p^*$ measured in the original graph is upper bounded by
\begin{equation}\label{eqn:dis_Hamming_multiscale}\small
\begin{split}
\Ep&\l[D_H(\p^*,\hat{\p})\r]\le \min_{\delta\in[0,1]}\delta T \\
&+\sum_{\forall \hat{\Pd} \text{ s.t. } |\Delta(\hat{\Pd})|> \delta T }\l[|\Delta(\hat{\Pd})|\prod_{t\in \Delta(\hat{\Pd})}\theta\l(\frac{\mu}{2\sigma^2},|\hat{\V}_t|\r)\r]\\
&+\sum_{\forall \hat{\Pd}} f(T-|\Delta(\hat{\Pd})|)\prod_{t\in \Delta(\hat{\Pd})}\theta\l(\frac{\mu}{2\sigma^2},|\hat{\V}_t|\r),
\end{split}
\end{equation}
where $\forall \hat{\Pd}$ means $\text{all possible paths $\hat{\Pd}=(\hat{\V}_1,\hat{\V}_2, \ldots \hat{\V}_T)$}$ in the super-graph, $\Delta(\hat{\Pd})\subset\{1,2, \ldots T\}$ is the set of time indices $t$ at which $\hat{\V}_t\neq\V^*_t$, and $\theta(\cdot,\cdot)$ is defined in \eqref{eqn:funcf}.}
\end{theorem}

\begin{IEEEproof}
We partition all possible paths $\hat{\p}=(\hat{v}_1,\ldots \hat{v}_T)$ into groups, so that all paths in the group $\M_{\hat{\Pd}}$ have the same projected path $\hat{\Pd}$ in the super-graph. Since there are $m$ clusters in total, we have at most $m^T$ groups. For each coarse path $\hat{\Pd}=(\hat{\V}_1,\ldots \hat{\V}_T)$, denote by $\Delta(\hat{\Pd})\subset\{1,2, \ldots T\}$ the set of time indices $t$ at which $\hat{\V}_t\neq\V^*_t$. For each path $\hat{\p}=(\hat{v}_1,\ldots \hat{v}_T)$, denote by $L(\hat{\p})\subset\{1,2, \ldots T\}$ the set of time indices $t$ at which $\hat{v}_t\neq v^*_t$.

Now for each group $\M_{\hat{\Pd}}$, if one path in this group is chosen as the final path estimate $\hat{\p}$, two things must happen. The first is that the sum signal on the projected path $\hat{\Pd}$ is larger than the sum signal on the true projected path $\Pd^*$, because only in this way can we choose $\hat{\Pd}$ as the coarse path estimate when calling Algorithm~\ref{alg:dp_app} in the INITIALIZE step of the multiscale Viterbi decoding Algorithm~\ref{alg:dp_multi_scale}. This event is equivalent to
\begin{equation}\small
\sum_{t\in \Delta(\hat{\Pd})}\u_t(\hat{\V}_t)\ge \sum_{t\in \Delta(\hat{\Pd})}\u_t(\V^*_t).
\end{equation}
The second thing that must happen is that $\hat{\p}$ is just the set of nodes that achieves the maximum signals in $\hat{\Pd}=(\hat{\V}_1,\hat{\V}_2\ldots \hat{\V}_T)$. In particular, for $t\in L(\hat{\p})\setminus \Delta(\hat{\Pd})$, which is the set of time indices at which the coarse path estimate $\hat{\Pd}$ and the true coarse path $\Pd^*$ overlap but the path estimate $\hat{\p}$ and the true path $\p^*$ differ, we must have
\begin{equation}\small
\y_t(\hat{v}_t)\ge \y_t(v^*_t),\forall t\in L(\hat{\p})\setminus \Delta(\hat{\Pd})
\end{equation}
These two events are independent of each other, because the set of time points that they involve do not overlap, and all observations in different time points are independent of each other. Define $W_t=\u_t(\hat{\V}_t)$ and $U_t=\u_t(\V^*_t)$. Define $\beta_t=\y_t(\hat{v}_t)$ and $\alpha_t=\y_t(v^*_t)$.

The Hamming distance between the path estimate $\hat{\p}$ and the true path $\p^*$ is the cardinality of the set $L(\hat{\p})$, which is equal to the summation $|L(\hat{\p})\setminus \Delta(\hat{\Pd})|+|\Delta(\hat{\Pd})|$. Using the union bound, the expectation of the overall Hamming distance between the path estimate
$\hat{\p}$ and the true path $\p^*$ can thus be upper-bounded by
\begin{equation}\label{eqn:Hamming_split}\small
\begin{split}
\Ep[D_H(\p^*,\hat{\p})]&=\Ep[|\Delta(\hat{\Pd})|]+\Ep[|L(\hat{\p})\setminus \Delta(\hat{\Pd})|].
\end{split}
\end{equation}

Now we look at the first part. Using the union bound
\begin{equation}\small
\begin{split}
  &\Ep[|\Delta(\hat{\Pd})|]
\le \sum_{\forall\hat{\Pd}} |\Delta(\hat{\Pd})|\Pr\l( \sum_{t\in \Delta(\hat{\Pd})}\u_t(\hat{\V}_t)\ge \sum_{t\in \Delta(\hat{\Pd})}\u_t(\V^*_t)\r).
\end{split}
\end{equation}
First, we upper-bound the term $p_1:=\Pr\l( \sum_{t\in \Delta(\hat{\Pd})}\u_t(\hat{\V}_t)\ge \sum_{t\in \Delta(\hat{\Pd})}\u_t(\V^*_t)\r)$. Notice that the event $\sum_{t\in \Delta(\hat{\Pd})}\u_t(\hat{\V}_t)\ge \sum_{t\in \Delta(\hat{\Pd})}\u_t(\V^*_t)$ is the same as the event $\sum_{t=1}^T\u_t(\V^*_t)\le\sum_{t=1}^T\u_t(\hat{\V}_t)$, because $\Delta(\hat{\Pd})$ is the set of time indices $t$ at which $\V^*_t\neq \hat{\V}_t$. This is just the event that the sum signal on $\hat{\Pd}$ is greater that on $\Pd^*$. By Lemma~\ref{lmm:pro_upb},
%
%
%
\begin{equation}\label{eqn:upbp1}\small
p_1\le\prod_{t\in \Delta}\theta\l(\frac{\mu}{2\sigma^2},|\hat{\V}_t|\r),
\end{equation}
where $\theta(\cdot,\cdot)$ is defined in \eqref{eqn:funcf}. Therefore,
\begin{equation}\small
\Ep[|\Delta(\hat{\Pd})|]\le \sum_{\forall\hat{\Pd}} |\Delta(\hat{\Pd})|\prod_{t\in \Delta(\hat{\Pd})}\theta\l(\frac{\mu}{2\sigma^2},|\hat{\V}_t|\r).
\end{equation}
If we only compute the sum over the paths such that $|\Delta(\hat\Pd)|>\delta T$, we can upper bound the expected Hamming distances of other paths by $\delta T$. Therefore,
\begin{equation}\label{eqn:up_der1}\small
\Ep[|\Delta(\hat{\Pd})|]\le \delta T+\sum_{\forall\hat{\Pd} \text{ s.t. } |\Delta(\hat\Pd)|>\delta T} |\Delta(\hat{\Pd})|\prod_{t\in \Delta(\hat{\Pd})}\theta\l(\frac{\mu}{2\sigma^2},|\hat{\V}_t|\r).
\end{equation}

For the second term $\Ep[|L(\hat{\p})\setminus \Delta(\hat{\Pd})|]$, we use the union bound for all paths in $\M_{\hat{\Pd}}$, i.e., for all paths whose projected path is $\hat{\Pd}$. Using the union bound, we have
\begin{equation}\small
\begin{split}
&\Ep[|L(\hat{\p})\setminus \Delta(\hat{\Pd})|]
\le \sum_{\forall\hat{\Pd}}\Pr\l( \sum_{t\in \Delta(\hat{\Pd})}\u_t(\hat{\V}_t)\ge \sum_{t\in \Delta(\hat{\Pd})}\u_t(\V^*_t)\r)\\
&\cdot\sum_{\hat{\p}\in \M_{\hat{\Pd}}}|L(\hat{\p})\setminus \Delta(\hat{\Pd})|\prod_{t\in L\setminus S}\Pr(\hat{v}_t\neq v^*_t)\prod_{t\in L^c}\Pr(\hat{v}_t=v^*_t),
\end{split}
\end{equation}
where the $\prod_{t\in L\setminus S}\Pr(\hat{v}_t\neq v^*_t)\prod_{t\in L^c}\Pr(\hat{v}_t=v^*_t)$ factor represents the probability that for $t\in L(\hat{\p})\setminus \Delta(\hat{\Pd})$, the coarse path estimate $\hat{\Pd}$ and the true coarse path $\Pd^*$ overlap but the path estimate $\hat{\p}$ and the true path $\p^*$ differ. Now look at the second line of the above inequality. For a fixed path $\hat{\p}$ with a fixed projected path estimate $\hat{\Pd}$,
\begin{equation}\small
|L(\hat{\p})\setminus \Delta(\hat{\Pd})|=\sum_{\tau\in[T]\setminus \Delta(\hat{\Pd})}\one_{(\hat{v}_\tau\neq v^*_\tau)}.
\end{equation}
By changing the order of summation,
\begin{equation}\small
\begin{split}
&\sum_{\hat{\p}\in \M_{\hat{\Pd}}}|L(\hat{\p})\setminus \Delta(\hat{\Pd})|\prod_{t\in L\setminus S}\Pr(\hat{v}_t\neq v^*_t)\prod_{t\in L^c}\Pr(\hat{v}_t=v^*_t)\\
=&\sum_{\hat{\p}\in \M_{\hat{\Pd}}}\sum_{\tau\in[T]\setminus \Delta(\hat{\Pd})}\one_{(\hat{v}_\tau\neq v^*_\tau)}\prod_{t\in L\setminus S}\Pr(\hat{v}_t\neq v^*_t)\prod_{t\in L^c}\Pr(\hat{v}_t=v^*_t)\\
=&\sum_{\tau\in[T]\setminus \Delta(\hat{\Pd})}\sum_{\hat{\p}\in \M_{\hat{\Pd}}}\one_{(\hat{v}_\tau\neq v^*_\tau)}\prod_{t\in L\setminus S}\Pr(\hat{v}_t\neq v^*_t)\prod_{t\in L^c}\Pr(\hat{v}_t=v^*_t)\\
=&\sum_{\tau\in[T]\setminus \Delta(\hat{\Pd})}\sum_{\hat{\p}\in \M_{\hat{\Pd}}\text{ s.t. }\hat{v}_\tau\neq v^*_\tau}\prod_{t\in L\setminus S}\Pr(\hat{v}_t\neq v^*_t)\prod_{t\in L^c}\Pr(\hat{v}_t=v^*_t).
\end{split}
\end{equation}
In the last summation, we sum up the binomial terms for all paths that satisfy $\hat{v}_\tau\neq v^*_\tau$, so the final result will only be $\Pr(\hat{v}_\tau\neq v^*_\tau)$. Therefore,
\begin{equation}\small
\begin{split}
&\sum_{\hat{\p}\in \M_{\hat{\Pd}}}|L(\hat{\p})\setminus \Delta(\hat{\Pd})|\prod_{t\in L\setminus S}\Pr(\hat{v}_t\neq v^*_t)\prod_{t\in L^c}\Pr(\hat{v}_t=v^*_t)\\
=&\sum_{\tau\in[T]\setminus \Delta(\hat{\Pd})}\Pr(\hat{v}_\tau\neq v^*_\tau)
=\sum_{t\in[T]\setminus \Delta(\hat{\Pd})}\Pr(\y_t(\hat{v}_t)\ge \y_t(v^*_t)).
\end{split}
\end{equation}
The last equality is because from the earlier discussion, for the time $t\in [T]\setminus \Delta(\hat{\Pd})$ (when the coarse projected path $\hat{\Pd}$ overlaps with the true path $\Pd^*$), the estimate $\hat{v}_t\neq v^*_t$ if and only if $\y_t(\hat{v}_t)\ge \y_t(v^*_t)$. Then, we upper-bound the term $p_2:=\Pr(\y_t(\hat{v}_t)\ge \y_t(v^*_t))=\Pr(\beta_t\ge \alpha_t)$. Using the Markov inequality, for all $\gamma>0$,
\begin{equation}\small
\begin{split}
  p_2&=\Pr\l(\exp\l(\gamma(\beta_t-\alpha_t)\r)\ge 1\r)\le
  \min_{\gamma>0}\Ep\l[\exp\l(\gamma(\beta_t-\alpha_t)\r) \r]\\
  &=\min_{\gamma>0}\Ep\l[e^{\gamma\beta_t}\r]\Ep\l[e^{-\gamma\alpha_t}\r].
\end{split}
\end{equation}

From the definitions of $\beta_t$, we know for $t\in L(\hat{\p})\setminus \Delta(\hat{\Pd})$, the estimated path $\hat{\Pd}$ and the true path $\Pd^*$ overlap, i.e., $\hat{\V}_t=\V^*_t$. This means $\beta_t$ is the maximum of $|\hat{\V}_t|-1=|\V^*_t|-1$ i.i.d. random variables $W_i^\text{off}$, where each $W_i^\text{off}\overset{\D}{=}W^\text{off}\sim\N(0,\sigma^2)$. 
From the definition of $\alpha_t$, $\alpha_t$ has the same distribution as $U^\text{on}\sim\N(\mu,\sigma^2)$. Therefore,
using the same large-deviation bounding techniques from the proof of Lemma~\ref{lmm:pro_upb} (see \eqref{eqn:ineq_off} to \eqref{eqn:fsl} for details), we have
\begin{equation}\label{eqn:upbp2}\small
\begin{split}
  p_2\le \theta\l(\frac{\mu}{2\sigma^2},|\V^*_t|\r).
\end{split}
\end{equation}
Finally
\begin{equation}\small
\begin{split}
&\Ep[|L(\hat{\p})\setminus \Delta(\hat{\Pd})|]\\
&\le \sum_{\forall\hat{\Pd}}\Pr\l( \sum_{t\in \Delta(\hat{\Pd})}\u_t(\hat{\V}_t)\ge \sum_{t\in \Delta(\hat{\Pd})}\u_t(\V^*_t)\r)\\
&\quad\sum_{\hat{\p}\in \M_{\hat{\Pd}}}|L(\hat{\p})\setminus \Delta(\hat{\Pd})|\prod_{t\in L\setminus S}\Pr(\hat{v}_t\neq v^*_t)\prod_{t\in L^c}\Pr(\hat{v}_t=v^*_t)\\
&\overset{(a)}{\le} \sum_{\forall\hat{\Pd}}\prod_{t\in \Delta(\hat{\Pd})}\theta\l(\frac{\mu}{2\sigma^2},|\hat{\V}_t|\r)\sum_{t\in[T]\setminus \Delta(\hat{\Pd})}\Pr(\y_t(\hat{v}_t)\ge \y_t(v^*_t))\\
&\overset{(b)}{\le}\sum_{\forall\hat{\Pd}}\prod_{t\in \Delta(\hat{\Pd})}\theta\l(\frac{\mu}{2\sigma^2},|\hat{\V}_t|\r)\sum_{t\in[T]\setminus \Delta(\hat{\Pd})}\theta\l(\frac{\mu}{2\sigma^2},|\V^*_t|\r),
\end{split}
\end{equation}
where (a) follows from \eqref{eqn:upbp1} and (b) follows from \eqref{eqn:upbp2}. Using the definition of the first-$k$ sum, we have
\begin{equation}\label{eqn:up_der2}\small
\Ep[|L(\hat{\p})\setminus \Delta(\hat{\Pd})|]\le \sum_{\forall\hat{\Pd}}f(T-|\Delta(\hat{\Pd})|)\prod_{t\in \Delta(\hat{\Pd})}\theta\l(\frac{\mu}{2\sigma^2},|\hat{\V}_t|\r).
\end{equation}
Plugging \eqref{eqn:up_der1} and \eqref{eqn:up_der2} into \eqref{eqn:Hamming_split}, we complete the proof.
\end{IEEEproof}

From Theorem~\ref{thm:destination}, the destination distance between the output $\hat{\p}$ of Algorithm~\ref{alg:dp_multi_scale} (a chain of nodes in the original graph) and the true path $\p^*$ can be trivially upper-bounded by replacing the distance $d(\V_T,\V^*_T)$ with the maximum possible distance between two nodes respectively in $\V_T$ and $\V^*_T$ (see the following theorem).

\begin{theorem}\label{thm:Destination_multiscale}
(Destination distance in Algorithm~\ref{alg:dp_multi_scale}) Let $\hat{\p}$ be the estimated chain of nodes using Algorithm~\ref{alg:dp_multi_scale} and let $\p^*$ be the true path. Let $\Pd^*$ be the projected path of $\p^*$. The expectation of the destination distance between $\hat{\p}$ and the true path $\p^*$ measured in the original graph is upper bounded by
\begin{equation}\label{eqn:dis_destination_multiscale}
\Ep\l[D_F(\p^*,\hat{\p})\r]\le \sum_{\forall \hat{\Pd}}\l[d_{\text{max}}(\V^*_T,\hat{\V}_T)\prod_{t\in \Delta(\hat{\Pd})}\theta\l(\frac{\mu}{2\sigma^2},|\hat{\V}_t|\r)\r],
\end{equation}
where $\forall \hat{\Pd}$ means $\text{all possible paths $\hat{\Pd}=(\hat{\V}_1,\hat{\V}_2, \ldots \hat{\V}_T)$}$, $\Delta(\hat{\Pd})\subset\{1,2, \ldots T\}$ is the set of time indices $t$ at which $\hat{\V}_t\neq\V^*_t$, $\theta(\cdot,\cdot)$ is defined in \eqref{eqn:funcf}, and $d_{\text{max}}(\V^*_T,\hat{\V}_T)$ is the maximum distance between two nodes in the two clusters $\V^*_T$ and $\hat{\V}_T$
\begin{equation}
d_{\text{max}}(\V^*_T,\hat{\V}_T)=\max_{v_T\in \V^*_T,\hat{v}_T\in \hat{\V}_T}d(v_T,\hat{v}_T).
\end{equation}
\end{theorem}

The bounds in Theorem~\ref{thm:Hamming} to Theorem~\ref{thm:Destination_multiscale} are of little use if we cannot compute them. Since there are an exponential number of possible paths in $T$ time points, one may think that the three bounds are not computable. However, the special structure of the two bounds (a sum-product structure) makes them computable in polynomial time.
\begin{theorem}\label{thm:comp_time}
The upper bound on the expected Hamming distance between the chain estimate $\hat{\p}$ and the true path in~\eqref{eqn:dis_Hamming} can be computed in time $\O(mT^2)$, while the upper bound on the expected destination distance in~\eqref{eqn:dis_destination} can be computed in time $\O(mT)$, where $m$ is the number of nodes in the super-graph $\G^\text{new}=(\V^\text{new},\E^\text{new})$, and $T$ is the number of time points. The upper bound on the expected Hamming distance in~\eqref{eqn:dis_Hamming_multiscale} can be computed in time $\O(mT^2)$, and the upper bound on the expected Destination distance in \eqref{eqn:dis_destination_multiscale} can also be computed in time $\O(mT)$.
\end{theorem}
\begin{IEEEproof}
We will only look at the two bounds in Theorem~\ref{thm:destination} and Theorem \ref{thm:Hamming_multiscale}, because the bound in Theorem~\ref{thm:Hamming} is part of the bound in Theorem~\ref{thm:Hamming_multiscale}, and the bound in Theorem~\ref{thm:Destination_multiscale} has the same form as the bound in Theorem~\ref{thm:destination}.

For Theorem~\ref{thm:destination}, the expression to be computed is the RHS of
\begin{equation}\label{eqn:dis_destination_proof}\small
\Ep\l[D_F(\Pd^*,\hat{\Pd})\r]\le \sum_{\forall\hat{\Pd}}\l[d(\V^*_T,\hat{\V}_T)\prod_{t\in \Delta(\hat{\Pd})}\theta\l(\frac{\mu}{2\sigma^2},|\hat{\V}_t|\r)\r],
\end{equation}
where $\Delta(\hat{\Pd})\subset [T]$ is the set of time indices at which the coarse path estimate $\hat{\Pd}$ and the true coarse path $\Pd^*$ do not overlap. The term $d(\V^*_T,\hat{\V}_t)$ is the distance between the two destinations of the true path $\Pd^*$ and the estimate $\hat{\Pd}$. Now we show how to use a dynamic programming method to compute the RHS of \eqref{eqn:dis_destination_proof}. Define a \emph{subpath} of length $\tau$ of a path $\hat{\Pd}=(\hat{\V}_1,\hat{\V}_2, \ldots \hat{\V}_T)$ to be the path $\hat{\p}_\tau=(\hat{\V}_1,\hat{\V}_2, \ldots \hat{\V}_\tau)$. For a subpath $\hat{\Pd}_\tau$, define the set $\Delta(\hat{\Pd}_\tau)\subset [\tau]$ to be the set of time indices smaller or equal to $\tau$ at which $\hat{\Pd}_\tau$ and $\Pd^*_\tau$ (the subpath of length $\tau$ of the true coarse path $\Pd^*$) do not overlap. Define the \emph{partial sum} of order $\tau$ at node $\hat{V}_\tau$ as
\begin{equation}\label{eqn:upb_dp_destination}\small
S_\tau(\hat{\V}_\tau)=\sum_{\text{All possible subpaths $\hat{\Pd}_\tau$ that ends in $\hat{\V}_\tau$}} \prod_{t\in \Delta(\hat{\Pd}_\tau)} \theta\l(\frac{\mu}{2\sigma^2},|\hat{\V}_t|\r).
\end{equation}
The RHS of \eqref{eqn:dis_destination_proof} can be written as $\sum_{\V_T\in[m]}S_T(\hat{\V}_T)$. Our goal is to compute $S_T(\hat{\V}_T)$ for all $\hat{\V}_T\in [m]$ by inductively computing $S_\tau(\hat{\V}_\tau)$ for $\tau =1,2,\ldots T$ and all possible $\hat{\V}_\tau\in [m]$. First note that $S_1(\hat{\V}_1)$ for all $\hat{\V}_1\in [m]$ are quite easy to compute, because
\begin{equation}\small
S_1(\hat{\V}_1)=\left\{
                \begin{array}{cc}
                  \theta\l(\frac{\mu}{2\sigma^2},|\hat{\V}_t|\r), & \text{if $\hat{\V}_1\neq \V^*_1$},\\
                  1, & \text{if $\hat{\V}_1=\V^*_1$}.
                \end{array}
              \right.
\end{equation}
where recall that $\V^*_1$ is the starting point of the true coarse path $\Pd^*=(\V^*_1,\ldots \V^*_T)$. For $\tau>1$, we use the induction
\begin{equation}\label{eqn:dp_subpath_destination}\small
\begin{split}
&S_\tau(\hat{\V}_\tau)\\
&=\left\{
                \begin{array}{cc}
                  \theta\l(\frac{\mu}{2\sigma^2},|\hat{\V}_\tau|\r)\sum_{\hat{\V}_{\tau-1}\in\N(\hat{\V}_\tau)}S_{\tau-1}(\hat{\V}_{\tau-1}), & \text{if $\hat{\V}_\tau\neq \V^*_\tau$},\\
                  \sum_{\hat{\V}_{\tau-1}\in\N(\hat{\V}_\tau)}S_{\tau-1}(\hat{\V}_{\tau-1}), & \text{if $\hat{\V}_\tau=\V^*_\tau$}.
                \end{array}
              \right.
\end{split}
\end{equation}
The summation of $S_{\tau-1}(\hat{\V}_{\tau-1})$ is over the neighborhood of $\hat{\V}_\tau$, because any subpath $\hat{\Pd}_\tau$ that ends in $\hat{\V}_\tau$ can be viewed as a subpath $\hat{\Pd}_{\tau-1}$ that ends in the neighborhood of $\hat{\V}_\tau$ concatenated by $\hat{\V}_\tau$. The difference between the two cases in \eqref{eqn:dp_subpath_destination} is because when $\hat{\V}_\tau=\V^*_\tau$, $\tau$ is not in $\Delta(\hat{\Pd}_\tau)$ (recall that $\Delta(\hat{\Pd}_\tau)$ is the time when the subpath $\hat{\Pd}_\tau$ does not overlap with the true path $\Pd^*_\tau$) and the product $\prod_{t\in \Delta(\hat{\Pd}_\tau)} \theta\l(\frac{\mu}{2\sigma^2},|\hat{\V}_t|\r)$ in \eqref{eqn:dis_destination_proof} does not include $t=\tau$. Using this method, we can compute $S_T(\hat{\V}_T)$ for all $\hat{\V}_T\in [m]$ in $\O(mT)$ time.

For Theorem~\ref{thm:Hamming_multiscale}, the expression to be computed can be written as the RHS of \eqref{eqn:dis_Hamming_proof}
\begin{equation}\label{eqn:dis_Hamming_proof}\small
\begin{split}
\Ep\l[D_H(\p^*,\hat{\p})\r]\le& \sum_{\forall\hat{\Pd}} g(|\Delta(\hat{\Pd})|)\prod_{t\in \Delta(\hat{\Pd})}\theta\l(\frac{\mu}{2\sigma^2},|\hat{\V}_t|\r),
\end{split}
\end{equation}
where $g(|\Delta(\hat{\Pd})|)=\mathbf{1}_{\{|\Delta(\hat{\Pd})|>\delta T\}}|\Delta(\hat{\Pd})|+f(T-|\Delta(\hat{\Pd})|)$ can be evaluated for all possible values of $|\Delta(\hat{\Pd})|\in [T]$, because the first-k sum $f(k)$ for all $k$ can be directly computed by sorting the $T$ terms $\theta\l(\frac{\mu}{2\sigma^2},|\V^*_t|\r),t=1,2,\ldots T$, which has a negligible cost of $\mathcal{O}(T\log T)$. We still use the above method of induction using subpaths. However, when computing the RHS of \eqref{eqn:dis_Hamming_proof}, we need two variables for each subpath $\hat{\Pd}_\tau$: the last node $\hat{\V}_\tau$ on the path $\hat{\Pd}_\tau$ and the Hamming distance $|\Delta(\hat{\Pd}_\tau)|$ between the subpath $\hat{\Pd}_\tau$ and the true subpath $\Pd^*_\tau$. Therefore, we define the following partial sum of order $(\tau,w)$ at node $\hat{\V}_\tau$:
\begin{equation}\label{eqn:upb_dp_Hamming}\small
\begin{split}
  S_{\tau,w}(\hat{\V}_\tau)=\sum_{\substack{\text{All possible subpaths $\hat{\Pd}_\tau$ that ends}\\ \text{in $\hat{\V}_\tau$ such that $|\Delta(\hat{\Pd}_\tau)|=w$}}} \quad\prod_{t\in \Delta(\hat{\Pd}_\tau)} \theta\l(\frac{\mu}{2\sigma^2},|\hat{\V}_t|\r).
\end{split}
\end{equation}
Then, the RHS of \eqref{eqn:dis_Hamming_proof} can be written as $\sum_{\V_T\in[m]}\sum_{w\in[T]}g(w)S_{T,w}(\hat{\V}_T)$. Our goal is to compute $S_{T,w}(\hat{\V}_T)$ for all $\hat{\V}_T\in [m]$ and all $w=1\ldots T$ by inductively computing $S_{\tau,w}(\hat{\V}_\tau)$ for all $\tau =1,2,\ldots T$, all $w=0,1,\ldots T$ and all possible $\hat{\V}_\tau\in [m]$. We have to compute all partial sums for $w=0,1,\ldots T$ because the Hamming distance between two paths $\hat{\Pd}$ and $\Pd^*$ can be at most $T$. First note that $S_{1,w}(\hat{\V}_1)$ for all $\hat{\V}_1\in [m]$ are quite easy to compute, because
\begin{equation}\small
S_{1,w}(\hat{\V}_1)=\left\{
                \begin{array}{cc}
                  \theta\l(\frac{\mu}{2\sigma^2},|\hat{\V}_1|\r), & \text{if $\hat{\V}_1\neq \V^*_1$ and $w=1$},\\
                  1, & \text{if $\hat{\V}_1=\V^*_1$ and $w=0$},\\
                  0, & \text{otherwise}.
                \end{array}
              \right.
\end{equation}
For $\tau>1$, we use the induction
\begin{equation}\label{eqn:dp_subpath_Hamming}\small
S_{\tau,w}(\hat{\V}_{\tau})
=\left\{
                \begin{array}{c}
                  \theta\l(\frac{\mu}{2\sigma^2},|\hat{\V}_\tau|\r)\sum_{\hat{\V}_{\tau-1}\in\N(\hat{\V}_\tau)}S_{\tau-1,w-1}(\hat{\V}_{\tau-1}), \\
                   \qquad\qquad\qquad\qquad\text{if $\hat{\V}_\tau\neq \V^*_\tau$ and $w\ge 1$},\\
                  \sum_{\hat{\V}_{\tau-1}\in\N(\hat{\V}_\tau)}S_{\tau-1,w}(\hat{\V}_{\tau-1}), \\
                   \qquad\qquad\qquad\qquad\text{if $\hat{\V}_\tau=\V^*_\tau$ and $w\ge 1$},\\
                  0,  \text{if $\hat{\V}_\tau\neq\V^*_\tau$ and $w=0$},\\
                  1,  \text{if $\hat{\V}_\tau=\V^*_\tau$ and $w=0$}.
                \end{array}
              \right.
\end{equation}
Note that the above induction is different from the one in \eqref{eqn:dp_subpath_destination} in two places. First, in the first two cases of \eqref{eqn:dp_subpath_Hamming} that resemble the two cases in \eqref{eqn:dp_subpath_destination}, we compute the sum of either $S_{\tau-1,w-1}(\hat{\V}_{\tau-1})$ or $S_{\tau-1,w}(\hat{\V}_{\tau-1})$, depending on whether $\hat{\V}_{\tau}=\V^*_\tau$. This is because when $\hat{\V}_{\tau}\neq\V^*_\tau$, the Hamming distance $|\Delta(\hat{\p}_\tau)|$ has to increase by one compared to the case when $\hat{\V}_{\tau}=\V^*_\tau$. Second, in the third case of \eqref{eqn:dp_subpath_Hamming}, we have to deal with the case when $w=0$. The case $\hat{\V}_\tau\neq\V^*_\tau$ and $w=0$ cannot happen because $\hat{\V}_\tau\neq\V^*_\tau$ implies $w\ge 1$. The case $\hat{\V}_\tau=\V^*_\tau$ and $w=0$ happens only if the subpath $\hat{\Pd}_\tau$ is exactly the same subpath $\Pd_\tau$, in which case the product $\prod_{t\in \Delta(\hat{\Pd}_\tau)} \theta\l(\frac{\mu}{2\sigma^2},|\hat{\V}_t|\r)$ in \eqref{eqn:upb_dp_Hamming} has void index set $\Delta(\hat{\Pd}_\tau)$ and should equal to 1. Using this method, we can compute $S_{T,w}(\hat{\V}_T)$ for all $\hat{\V}_{T}\in [m]$ and $w=1,2,\ldots T$ in $\O(mT^2)$ time.
\end{IEEEproof}

\begin{remark}
Viterbi decoding can be viewed as a special type of message-passing techniques that can guarantee optimality. The possibility of analyzing classical Viterbi decoding comes from the Markov timeline structure in the message-passing graph (i.e., conditioned on the current state, past and future are independent of each other). However, the network partitioning technique in Algorithm~\ref{alg:dp_multi_scale} breaks this Markov property, which makes the analysis much harder than classical Viterbi decoding analysis. It is interesting to see if the proposed network partitioning technique and its analysis can apply to some other message-passing techniques that guarantee convergence \cite{du2016convergence,du2017convergence}.
\end{remark}
\section{A Case Study on Random Geometric Graphs}\label{sec:random_geometric}

\begin{figure}
  \centering
  \includegraphics[scale=0.3]{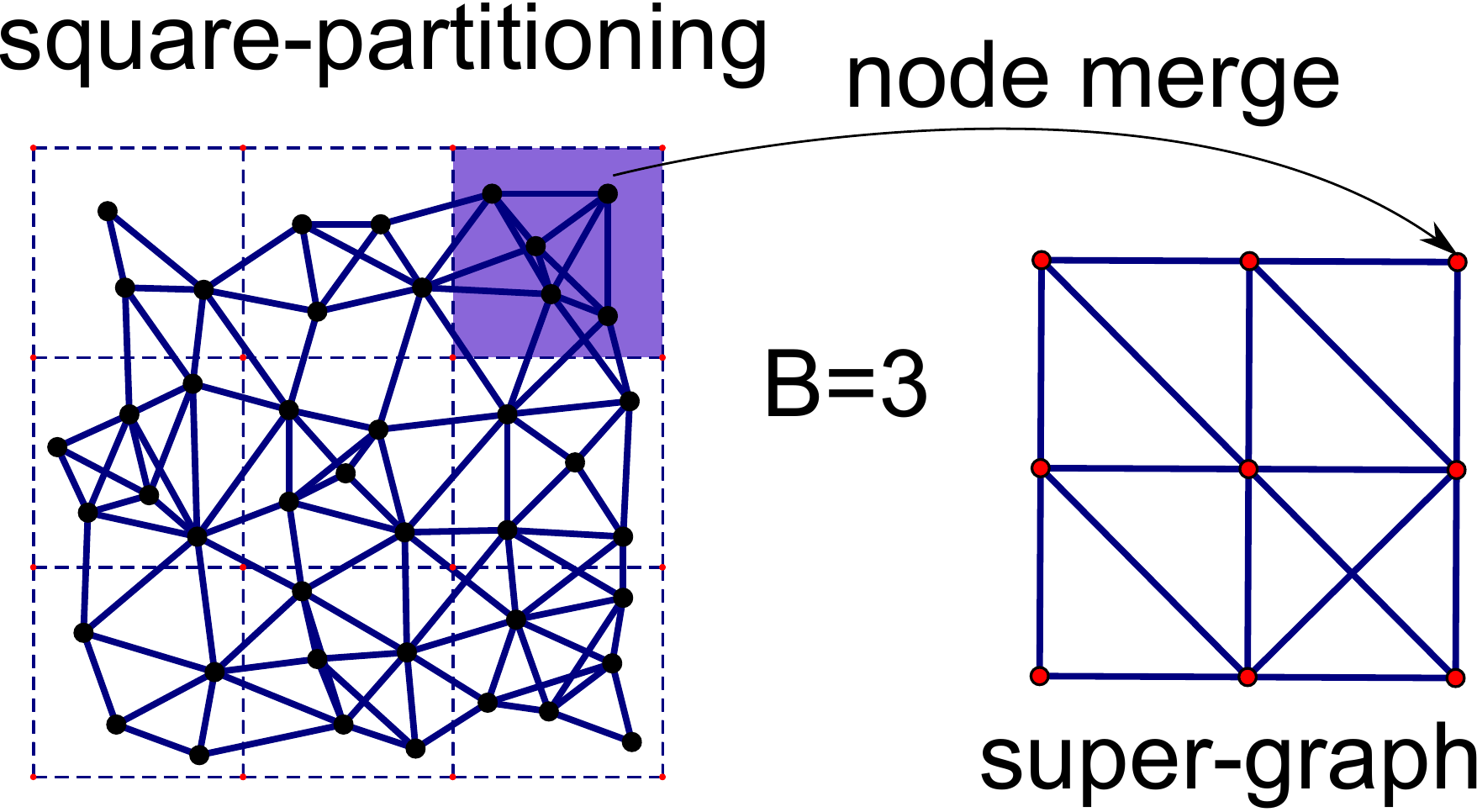}\\
  \caption{\textcolor{black}{Illustration of the square partitioning and the resulted super-graph when $B=3$.}}\label{fig:geo_ill}
\end{figure}

The random geometric graph is a good approximation to real sensor networks in the problem of tracking in a geographic area. Therefore, we study the analytical forms of the two bounds in Theorem~\ref{thm:Hamming} and Theorem~\ref{thm:destination} in the specific setting of a random geometric graph. The random geometric graph $\G=(\V,\E)$ that we use are composed of nodes that are distributed according to a Poisson point process with mean value $\lambda$ in a square area with length $1$. Two nodes are connected if they are within a threshold Euclidean distance $r\in (0,1)$. The partitioning of a random geometric graph can be done directly using a square partitioning.

\textcolor{black}{\begin{definition}\label{def:square_part}
(Square partitioning) Partition the square area of length $1$ into $B\times B$ congruent squares in which each square has length $\frac{1}{B}$. In this way, the node set $\V$ is also partitioned into $B^2$ clusters
\begin{equation}
\V=\bigcup_{j=1}^{B^2} \V_j,
\end{equation}
where each cluster, or super-node $\V_j$, corresponds to the set of nodes that are in the $j$-th square area of length $1/B$.
\end{definition}
We only consider the case when $1/B\ge r$, in which case two super-nodes $\V_i$ and $\V_j$ are connected in the super-graph $\G^\text{new}=(\V^\text{new},\E^\text{new})$ only if the two corresponding squares are adjacent (including diagonally adjacent). Therefore, the resulting super-graph of square partitioning is subgraph of a square lattice with diagonal connections (see Figure \ref{fig:geo_ill}).}

\subsection{Analysis of the Computational Complexity}
Denote by $\mathcal{C}$ the number of operations in the dynamic programming in the super-graph $\G^\text{new}$. Then, since the number of clusters is $B^2$, the computational complexity of Algorithm~\ref{alg:dp_app} is $\O(B^2T)$.

\subsection{Analysis of the Path-Localization Error}
\textcolor{black}{The result in Lemma~\ref{lmm:nodes_even} states that under the assumption of a Poisson point process, each square $\V_j$ contains approximately $\lambda\cdot\l( \frac{1}{B}\r)^2=\frac{\lambda}{B^2}$ nodes when $B^2=\O\l(\frac{n}{\log n}\r)$. This approximation can be formalized using the following lemma.
\begin{lemma}[\cite{toumpis2004large} Lemma 1]\label{lmm:nodes_even}
Suppose $B=\l(\lfloor\sqrt{\frac{n}{c_1\log n}}\rfloor\r)^2$. Then,
\begin{equation}\label{eqn:nodes_even}
  \Pr\l(\frac{c_1}{2}\log n\le|\V_j|\le 4c_1\log n,\forall j\r)>1-\frac{2n^{1-c_1/8}}{\log n}.
\end{equation}
\end{lemma}
Therefore, when the number of squares $B^2$ is not too large (when $B^2=\O\l(\frac{n}{\log n}\r)$), the number of the nodes in each square is approximately equal to each other (in the scaling sense) with high probability. Denote by $s_m$ the maximum number of nodes in one square. Then, we have $|\V_j|\le s_m$ for all super-nodes in the super-graph $\G^\text{new}$, and $s_m$ is approximately equal to $\frac{\lambda}{B^2}$ (in the scaling sense). Therefore, we can upper bound the RHS of \eqref{eqn:dis_Hamming} by replacing $|\hat{\V}_t|$ with $s_m$ without making the bound too loose, and obtain a bound in closed form at the same time.}
\begin{corollary}\label{cor:dis_Hamming}
In the random geometric graph $\G=(\V,E)$, the expectation of the Hamming distance between the path estimate and the true path measured on the super-graph is upper bounded by
\begin{equation}\label{eqn:dis_Hamming_2}
\Ep\l[D_H(\Pd^*,\hat{\Pd})\r]\le 9\exp\l(-\frac{\mu^2}{4\sigma^2}\r)s_m T,
\end{equation}
when $\mu/\sigma>2\sqrt{\log(9s_m)}$.
\end{corollary}

\begin{IEEEproof}
See Supplementary Section \ref{app:Hamming}.
\end{IEEEproof}

Therefore, one can see that the expectation of the Hamming distance between the path and the true path measured on the super-graph has an upper bound that grows linearly with time $T$ on a random geometric graph. One may ask whether this linear growth of localization error can be outperformed by some other localization techniques. The following theorem claims that even if one has the access to all the available information $\y_t(v),\forall v\in\V,t=1,\ldots T$, he/she still cannot obtain sublinear growth of localization error with $T$ on the super-graph. Since the only available information that we use in the coarsened dynamic programming in Algorithm \ref{alg:dp_app} is a lossy version of all the available information, the localization error cannot grow sub-linearly with $T$ using other localization algorithms.

\begin{theorem}\label{thm:lower_bound}
Suppose one uses an arbitrary path-localization estimator $\hat{\Pd}(\cdot)$ with arguments $\y_t(v),\forall v\in\V,t=1,\ldots T$. Then, there exists a constant $\eta$ independent of $T$ such that, for all $T$ sufficiently large, the path-localization error measured using the Hamming distance on the super-graph satisfies
\begin{equation}\label{eqn:lower_bound}
\Ep\l[D_H(\Pd^*,\hat{\Pd})\r]\ge \eta T=\Omega(T).
\end{equation}
\end{theorem}
\begin{IEEEproof}
The path estimator can only perform better if more accurate information  is given. Now we choose to give the following information: we partition  the time range $[1,T]\cap \Za$ into $\lceil\delta T\rceil$ intervals, where each interval has $1/\delta$ times slots (we choose $\delta$ such that $1/\delta$ is an integer):
\begin{equation}
[1,T]\cap \Za=\bigcup_{i=1}^{\lceil\delta T\rceil}(\frac{i-1}{\delta},\frac{i}{\delta}]\cap \Za.
\end{equation}
We choose the constant $\delta$ small enough so that the diameter (the maximum multi-hop distance between two nodes) of any cluster $\V_i$ in the graph partitioning $\V=\bigcup_{i=1}^m \V_i$ is smaller than $\frac{1}{2\delta}$. Now consider the path-localization problem with the side information on the exact positions of the moving agent at time points $t=\frac{1}{\delta},\frac{2}{\delta},\ldots\frac{\lceil\delta T\rceil}{\delta}$ on the true path $\p^*=(v^*_0,v^*_1,v^*_2,\ldots v^*_T)$ in the original graph. When this side information is provided, the path-localization problem is decomposed into $\lceil \delta T\rceil$ small path-localization subproblems with path length smaller or equal to $1/\delta$, because the localization in two consecutive subproblems are made independent by the fixed junction, i.e., the exact location at one of the time indices $t=\frac{1}{\delta},\frac{2}{\delta},\ldots\frac{\lceil\delta T\rceil}{\delta}$.

Now we look at the first path-localization problem with the end node $v^*_{1/\delta}$ given. Now that $\delta$ is small enough so that the diameter of any cluster $\V_i$ is smaller than $\frac{1}{2\delta}$, the projected path of the true path $(v^*_1,v^*_2,\ldots v^*_{1/\delta})$ in the super-graph is not necessarily a constant path (the path that stays at the same node). This means that the path localization on this time segment $t=1,2,\ldots 1/\delta$ is not trivial, i.e., one cannot directly assign the projected position of $v^*_{1/\delta}$ to the other time indices $t=1,2,\ldots 1/\delta-1$. Therefore, for any path-localization estimator $\hat{\Pd}$, the expected error of $\hat{\Pd}$ on this subproblem with path length $1/\delta$ cannot be zero. Denote by $\psi$ the expected path-localization error on this subproblem. Then, we know that the overall path-localization error is at least the summation of the path-localization error on each small problem. In other words,
\begin{equation}
\Ep\l[D_H(\Pd^*,\hat{\Pd})\r]\ge c\cdot \lceil\delta T\rceil=:\eta T.
\end{equation}
\end{IEEEproof}

\subsection{Simulation}

\begin{figure*}
\centering
\begin{subfigure}{.3\textwidth}
  \centering
  \includegraphics[scale=0.4]{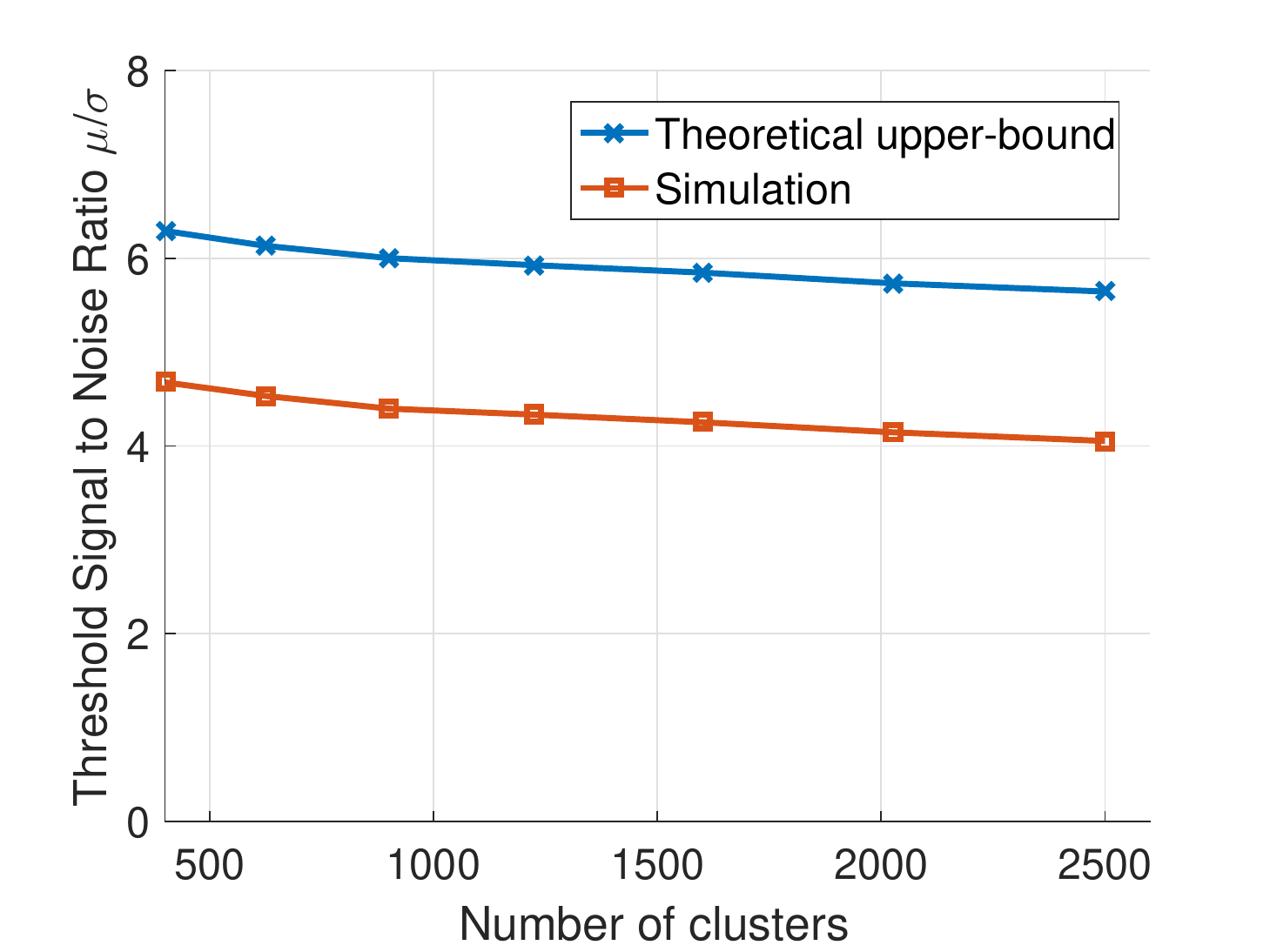}
  \caption{Threshold SNR for Hamming distance}
  \label{fig:geo_simu_Hamming}
\end{subfigure}%
\begin{subfigure}{.3\textwidth}
  \centering
  \includegraphics[scale=0.4]{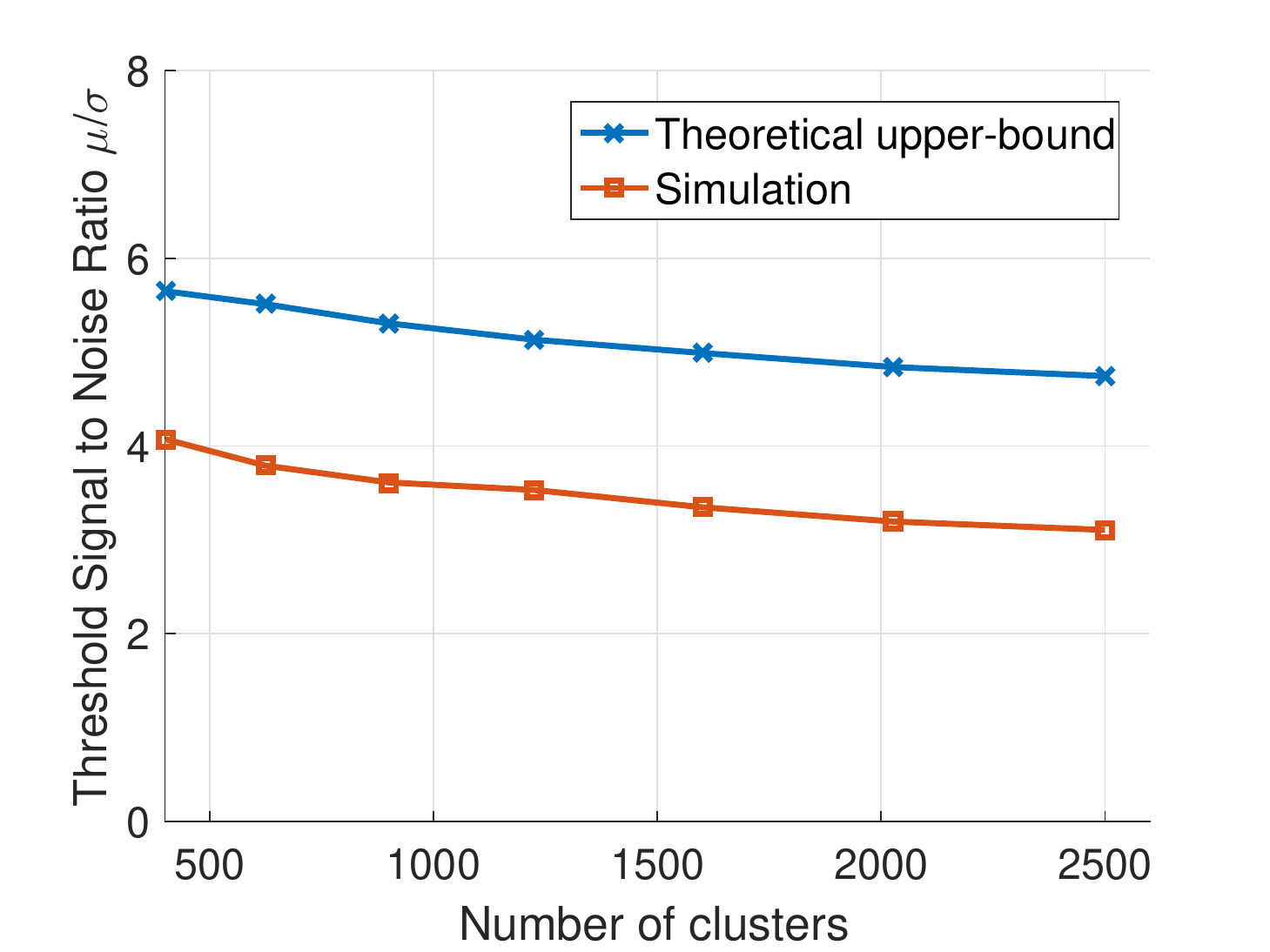}
  \caption{Threshold SNR for destination distance}
  \label{fig:geo_simu}
\end{subfigure}
\begin{subfigure}{.3\textwidth}
  \centering
  \includegraphics[scale=0.4]{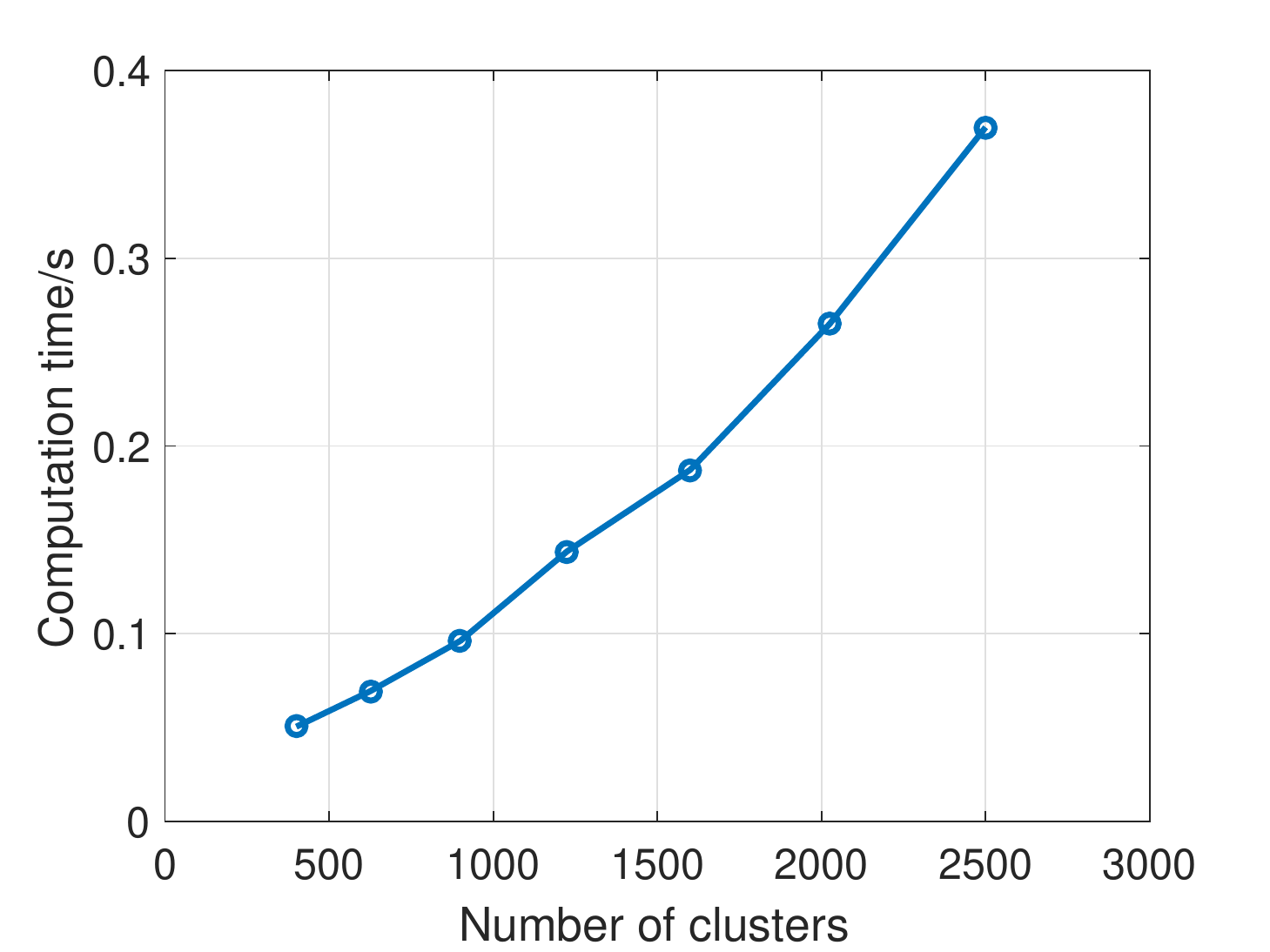}
  \caption{Computational complexity}
  \label{fig:geo_time}
\end{subfigure}
\caption{Figure (a) and (b): Threshold SNR to achieve 0.05 Hamming distance and 0.01 destination distance for different number of clusters (super-nodes). Figure (c): Computation time of one step in Algorithm~\ref{alg:dp_app} versus the number of clusters in the graph partitioning.}
\label{fig:geo_simulation}
\end{figure*}

First, we test the algorithm on a random geometric graph with 20000 randomly generated nodes that are distributed according to the Poisson point process on a square area with length 1. Two nodes are connected if they are within distance 0.02. Then, we partition the square areas into $m$ sub-squares using direct square tessellation and merge the nodes in each square into one ``super-node''. The number of clusters can be $m=400$, $625$, $900$, $1225$, $1600$, $2025$ or $2500$. After that, we generate a random walk on the graph to represent the positions of a moving agent and use Algorithm~\ref{alg:dp_multi_scale} to estimate both the trajectory and the final position of the path-signal from Gaussian observation noise. The destination distance error metric $d(\cdot,\cdot)$ in Definition~\ref{def:destination} is defined as the Euclidean distance on the square area.

The results in Fig.~\ref{fig:geo_simu_Hamming} and Fig.~\ref{fig:geo_simu} respectively show the threshold signal to noise ratio $\mu/\sigma$ to achieve Hamming localization error $\le 0.05$ and destination-localization error $\le 0.01$ versus different number of clusters in the graph partitioning stage. The theoretical upper-bounds on the required SNR are respectively obtained from Theorem~\ref{thm:Hamming_multiscale} and Theorem~\ref{thm:Destination_multiscale} by setting the desired Hamming distance to 0.05 and the desired destination distance to 0.01. The result in Fig.~\ref{fig:geo_time} shows the computation time of one step in the FOR-loop in Algorithm~\ref{alg:dp} when the number of clusters differ. We can see from Fig.~\ref{fig:geo_simu_Hamming} and Fig.~\ref{fig:geo_simu} that when the number of clusters increases, the required SNR to achieve the same localization error decreases, but the computation time increases. In practice, one should find the optimal number of clusters to obtain a tradeoff between computation time and localization error.

\section{A Case Study on a Real Graph}\label{sec:hub_graph}

\begin{figure*}
\centering
\begin{subfigure}{.3\textwidth}
  \centering
  \includegraphics[width=1\linewidth]{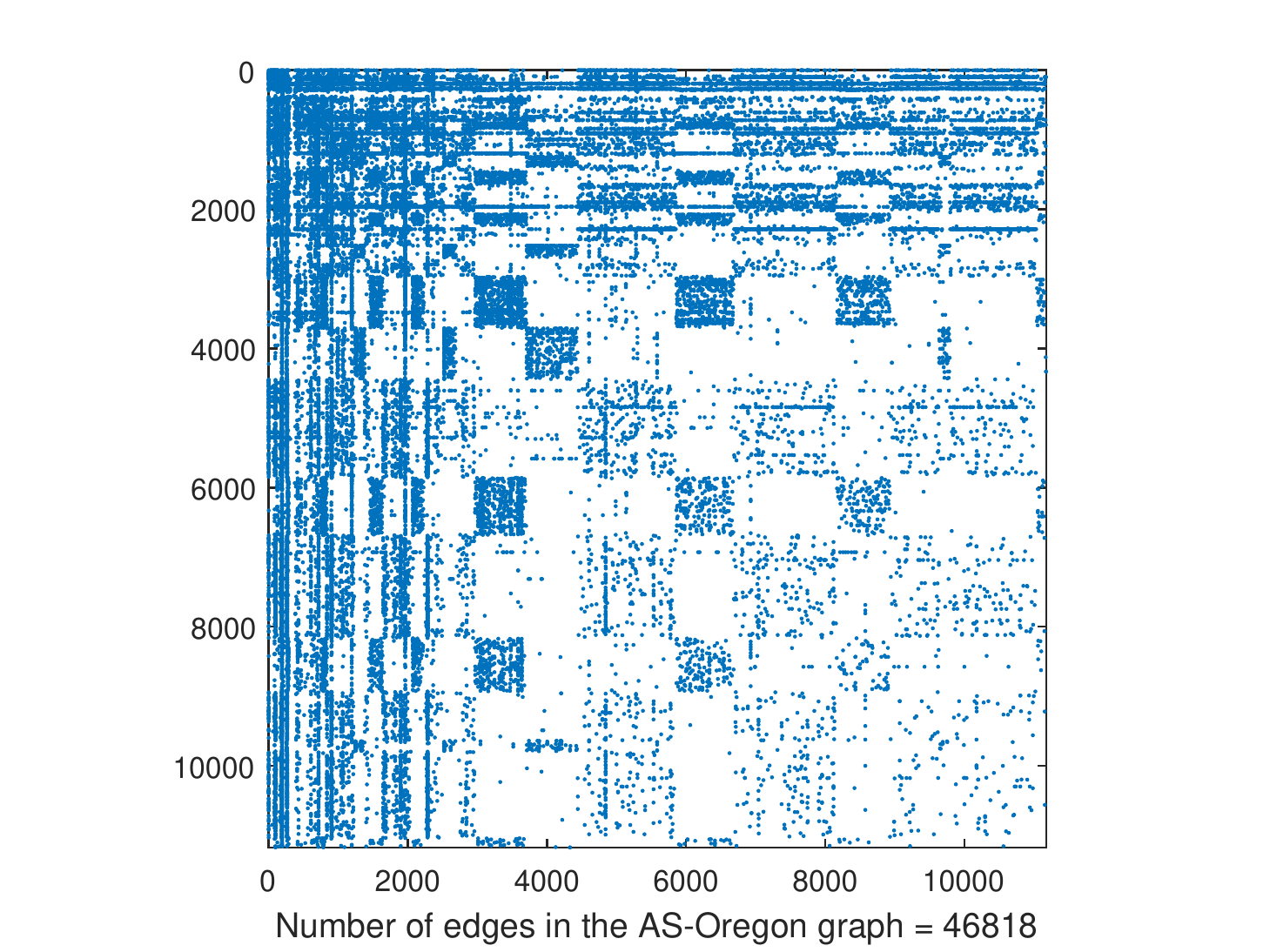}
  \caption{Adjacency matrix of AS-Oregon}
  \label{fig:sub1}
\end{subfigure}%
\begin{subfigure}{.3\textwidth}
  \centering
  \includegraphics[width=1\linewidth]{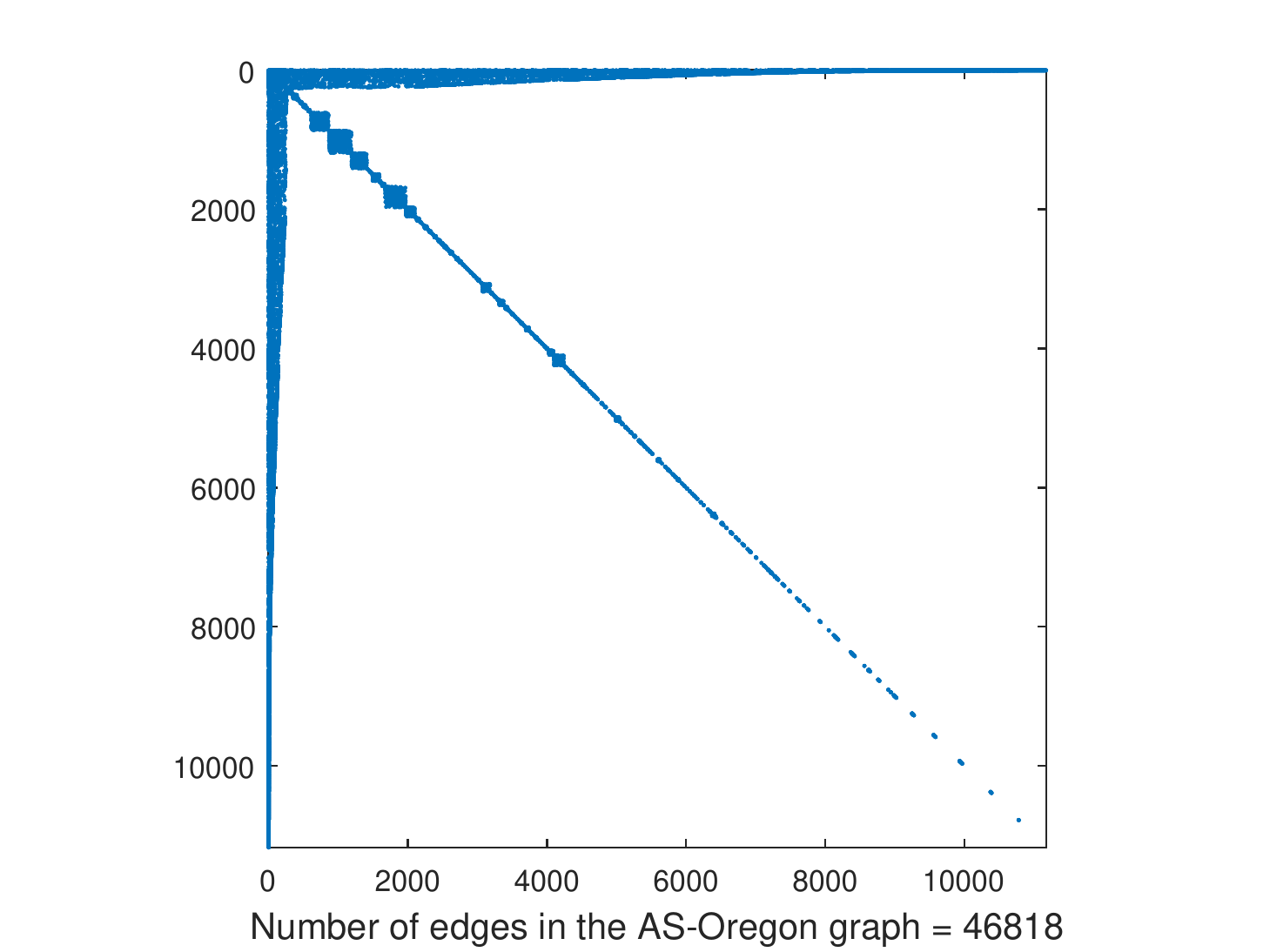}
  \caption{Adjacency matrix after Slashburn}
  \label{fig:sub2}
\end{subfigure}
\begin{subfigure}{.3\textwidth}
  \centering
  \includegraphics[width=1\linewidth]{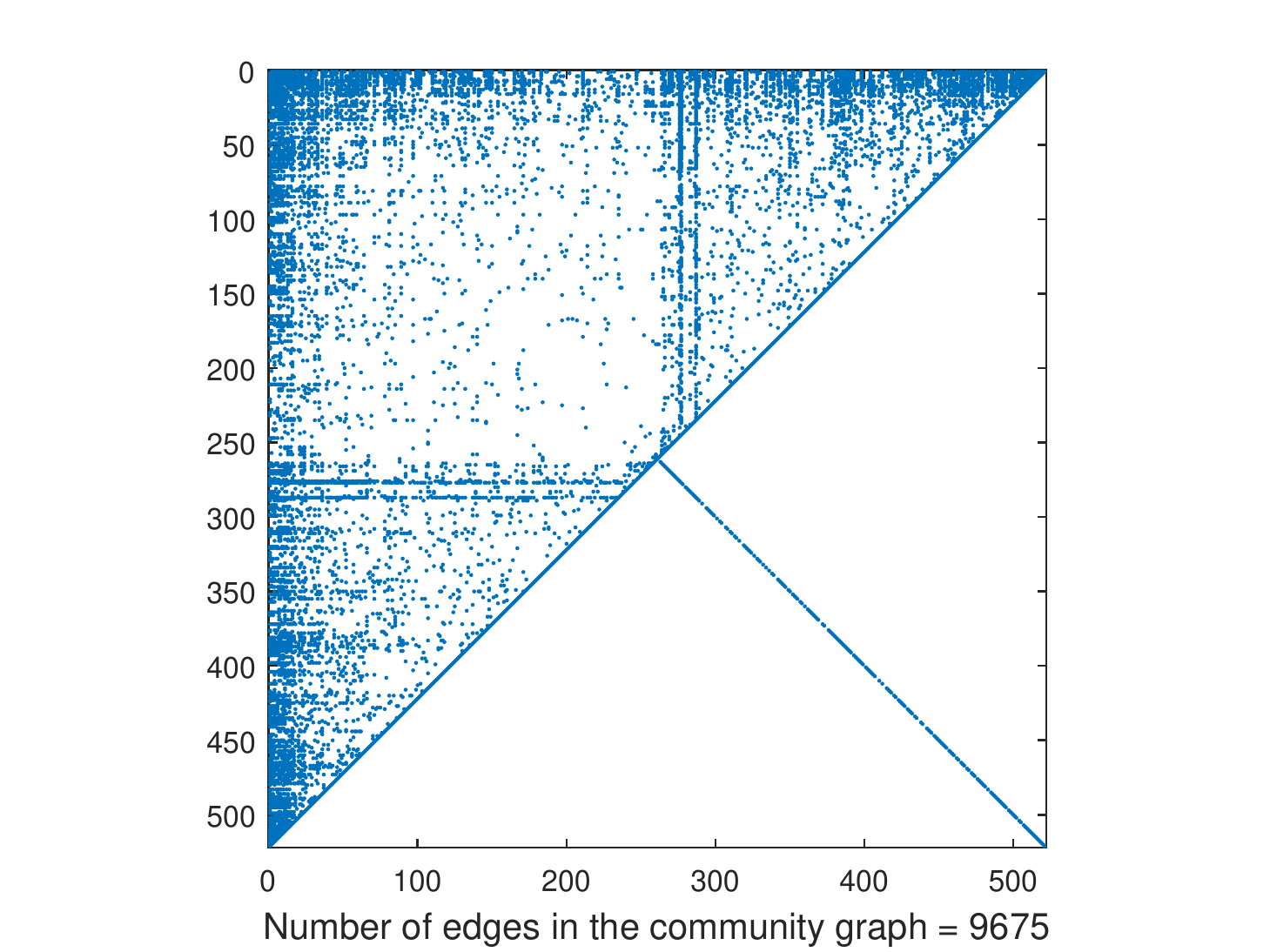}
  \caption{Adjacency matrix of the super-graph after Slashburn}
  \label{fig:sub3}
\end{subfigure}
\caption{An illustration of the Slashburn graph partitioning method}
\label{fig:AS_spy}
\end{figure*}

\textcolor{black}{We test different graph partitioning methods on a real graph. We focus on ``Slashburn'', which is a graph partitioning technique that can obtain a ``wing-shaped'' permuted adjacency matrix as shown in Fig. \ref{fig:sub2} from the original adjacency matrix in Fig. \ref{fig:sub1}. The main idea of \cite{lim2014slashburn} is that real-world graphs often do not have good cuts for a reasonable graph partitioning result, but can often be ``shattered'' into many small and unconnected clusters after a small set of hub-nodes are removed from the network. Fig. \ref{fig:AS_spy} is an example of Slashburn on the AS-Oregon graph \cite{leskovec2005graphs} of Autonomous Systems (AS) peering information inferred from Oregon route-views. In Fig. \ref{fig:sub2}, the hub-nodes are the ones on the upper-left corner of the adjacency matrix. After these hub-nodes are removed from the graph, the remaining nodes are scattered into many small clusters. For the partitioned graph in Fig. \ref{fig:sub2}, we define each connected component after removing the hub-nodes as one cluster, and define each hub-node as one cluster as well. Therefore, in the super-graph, we have one-node clusters formed by the hub-nodes. Then, we implement the multiscale Viterbi decoding algorithm as in Algorithm~\ref{alg:dp_multi_scale}.}

\textcolor{black}{Why do we use Slashburn for graph partitioning instead of other methods? This is because after we change the original graph into a super-graph, we wish the obtained super-graph is also sparse. In fact, from the obtained theoretical bound in Theorem~\ref{thm:Hamming_multiscale}, the localization error increases when the number of paths in the super-graph increases. Since the number of paths in the super-graph increases when the density of edges of the super-graph increases, we wish the obtained super-graph is sparse. Classical graph partitioning methods are not useful here because even if two clusters are connected by only one edge, these two clusters are connected in the super-graph. However, if we use the Slashburn method, the clusters are completely disconnected if hub-nodes are removed, which means the super-graph is still sparse because of these disconnected clusters (see the large blank space in Fig. \ref{fig:sub3}).}
\vspace{-2mm}
\subsection{Data experiments}
\vspace{-2mm}
\begin{figure}
  \centering
  \includegraphics[scale=0.4]{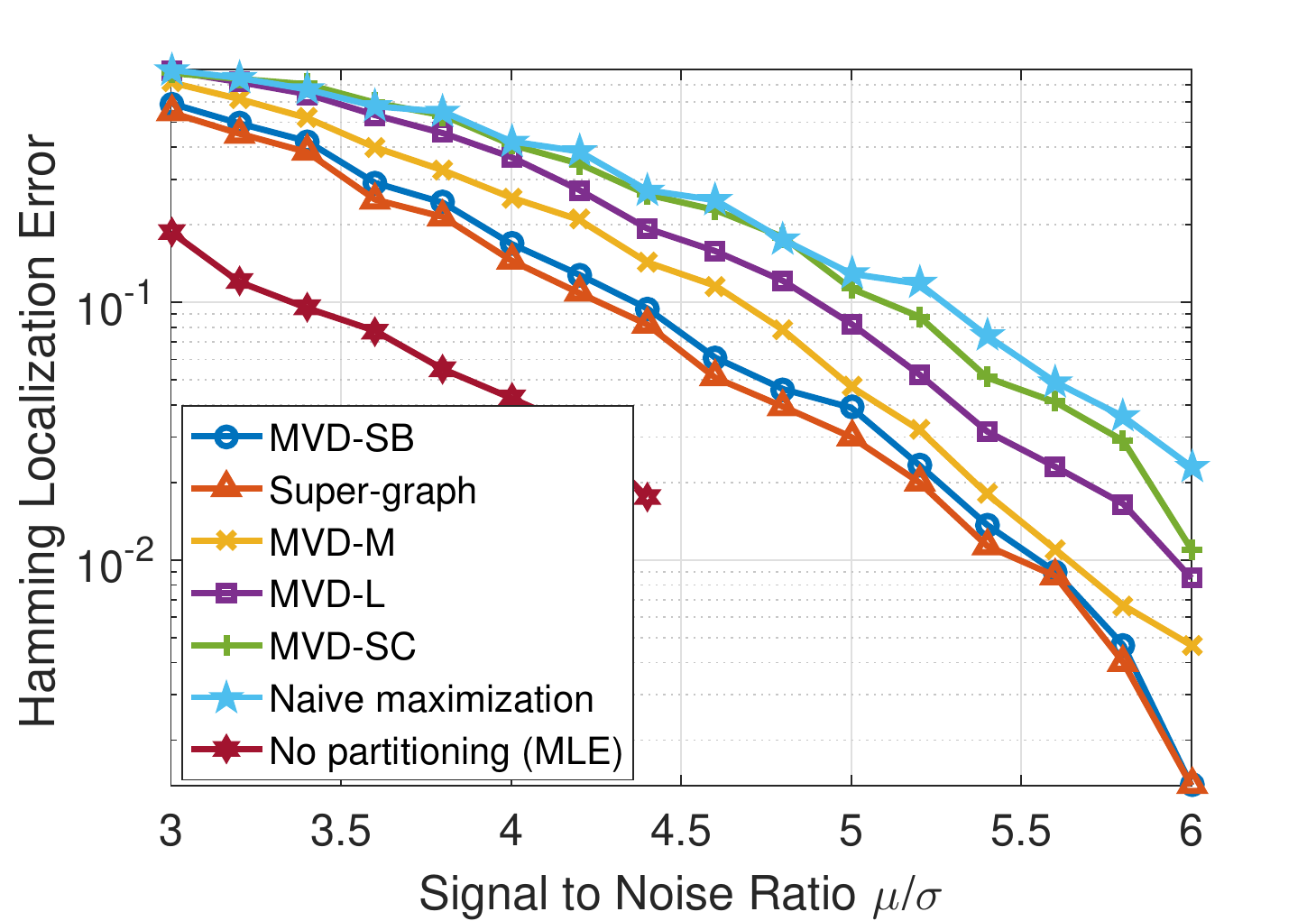}\\
  \caption{Localization error comparison between dynamic programming with and without graph partitioning on AS-Oregon graph. \textcolor{black}{The curve `no partitioning (MLE)' ends at $\mu/\sigma=4.5$ because the simulation without partitioning takes too long to obtain a steady and accurate simulation data point when the Hamming error is small.\vspace{-4mm}}}\label{fig:AS}
\end{figure}

\begin{figure}
  \centering
  \includegraphics[scale=0.4]{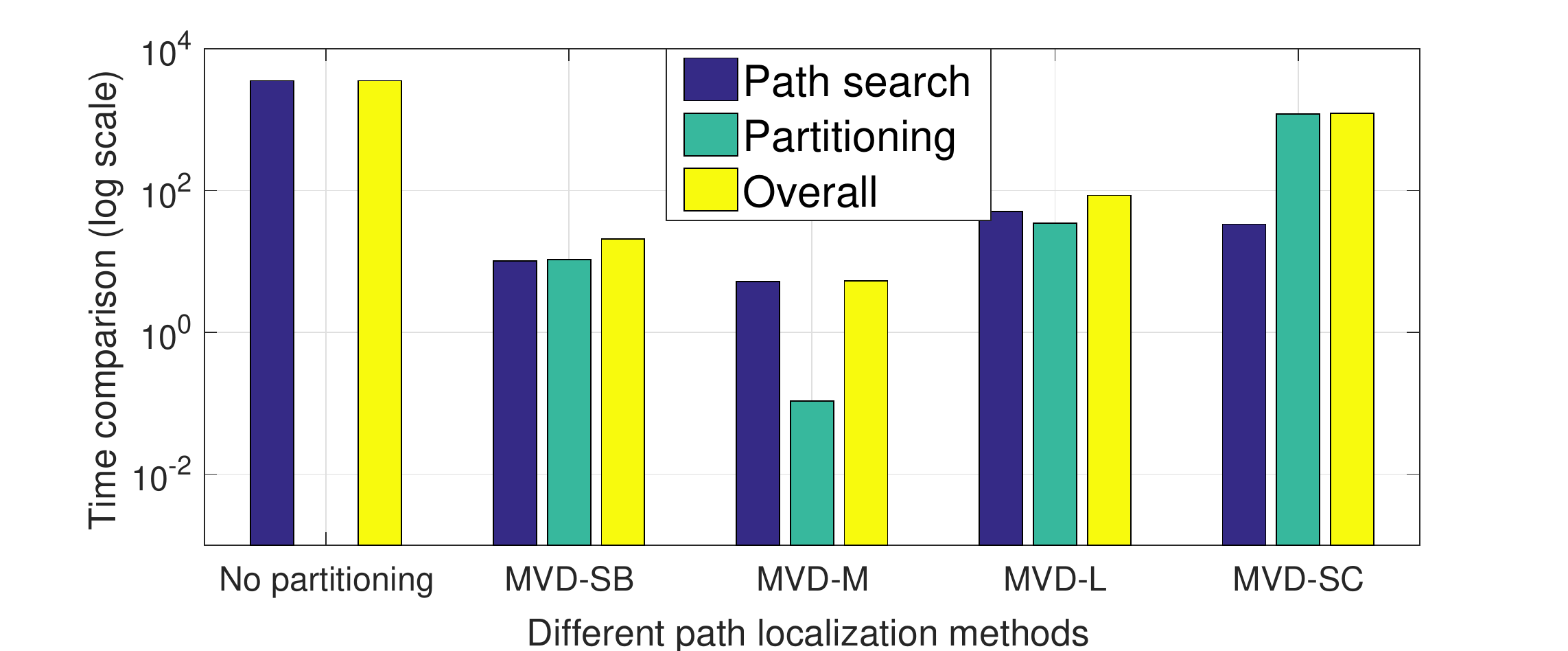}\\
  \caption{Time comparison between dynamic programming with and without graph partitioning on AS-Oregon graph. For each method, we show the time for path search in Algorithm~\ref{alg:dp_multi_scale} (blue), the time for graph partitioning (green), and the overall time (yellow).}\label{fig:AS_time}
\end{figure}

We test the multiscale Viterbi decoding algorithm on a real graph called AS-Oregon \cite{leskovec2005graphs}. The result on the localization error versus the signal to noise ratio $\mu/\sigma$ is shown in Fig.~\ref{fig:AS}. We also compare other partitioning methods including METIS \cite{karypis1998fast}, Louvain \cite{blondel2008fast} and spectral clustering \cite{von2007tutorial}. \textcolor{black}{The METIS algorithm is a two-phase algorithm, where in the first phase nodes are repeatedly merged based on recursive bipartite matchings and in the second phase the merged nodes are unfolded with local refinement. The Louvain algorithm is an iterative algorithm that seeks to maximizes the \emph{graph modularity} using local refinement.} The number of clusters in each method is:
\begin{table*}
  \begin{center}
    \begin{tabular}{ | l | l | l |l | l |}
    \hline
     & Slashburn & METIS & Louvain & Spectral Clustering \\ \hline
     Number of clusters & 521 & 500 & 1408 & 500 \\
    \hline
    \end{tabular}
\end{center}
\end{table*}
The number of clusters in the Slashburn method and the Lovain method is slightly higher because we cannot directly adjust the number of clusters.

The curves with legends ``MVD-SB'', ``MVD-M'', ``MVD-L'' and ``MVD-SC'' respectively refer to the proposed multiscale Viterbi decoding algorithm using the Slashburn, METIS, Louvain and Spectral Clustering. The curve with legend ``Naive maximization'' is the method that we mentioned in Section \ref{sec:sys_model} that chooses the node with the maximum node signal in the original graph at each time point. The curve with legend ``super-graph'' shows the localization error of Algorithm \ref{alg:dp_app} in the super-graph using the Slashburn algorithm. As we mentioned in Remark \ref{rmk:not_path_1}, one can see that the Hamming distance of Algorithm \ref{alg:dp_multi_scale} (the curve with legend ``MVD-SB'') is only slightly larger than that of Algorithm \ref{alg:dp_app} (the curve with legend ``super-graph''). This means that simply choosing the node with the maximum signal in each cluster in Algorithm \ref{alg:dp_multi_scale} is already near-optimal. Therefore, although one may use another round of dynamic programming to find the fine-grained path in the approximate path output by Algorithm \ref{alg:dp_app}, the obtained result has limited Hamming distance reduction compared to simply choosing the node with the maximum signal in each cluster.

The computation time of one step of dynamic programming with Slashburn (i.e., Algorithm~\ref{alg:dp_multi_scale}) and without graph partitioning (i.e., Algorithm~\ref{alg:dp} or the path MLE) are respectively 0.0101719 seconds and 3.5344 seconds. The partitioning time of Slashburn on the AS-Oregon graph is 10.5313 seconds. The number of time points is set to $T=1000$. Therefore, the total time of dynamic programming without partitioning is 3534.4 seconds, while the total time of the multiscale Viterbi decoding algorithm is 20.7032 seconds. The partitioning time can be further reduced if we tune a parameter that controls the hub-node size in Slashburn \cite{lim2014slashburn}, but the localization error gets higher. Similarly, if we reduce the size of each cluster, the localization error gets higher, but the computation time decreases. The computation time of dynamic programming with and without partitioning, including other partitioning methods, are also shown in Fig. \ref{fig:AS_time}.

Note that the destination distance (Euclidean distance) here does not have a specific meaning, so we only use the Hamming distance $D_H(\p^*,\hat{\p})=\sum_{t=1}^T\1(\hat{v}_t\neq v^*_t)$.

\bibliographystyle{ieeetr}
\bibliography{rough}

\begin{thebibliography}{10}

\bibitem{yang2017fast}
Y.~Yang, S.~Chen, M.~A. Maddah-Ali, P.~Grover, S.~Kar, and J.~Kovacevic, ``Fast
  path localization on graphs via multiscale viterbi decoding,'' in {\em IEEE
  International Conference on Acoustics, Speech and Signal Processing
  (ICASSP)}, pp.~4114--4118, IEEE, 2017.

\bibitem{ShumanNFOV:13}
D.~I. Shuman, S.~K. Narang, P.~Frossard, A.~Ortega, and P.~Vandergheynst, ``The
  emerging field of signal processing on graphs: {E}xtending high-dimensional
  data analysis to networks and other irregular domains,'' {\em IEEE Signal
  Process. Mag.}, vol.~30, pp.~83--98, May 2013.

\bibitem{SandryhailaM:14}
A.~Sandryhaila and J.~M.~F. Moura, ``Big data processing with signal processing
  on graphs,'' {\em IEEE Signal Process. Mag.}, vol.~31, no.~5, pp.~80--90,
  2014.

\bibitem{AnisGO:15}
A.~Anis, A.~Gadde, and A.~Ortega, ``Efficient sampling set selection for
  bandlimited graph signals using graph spectral proxies,'' {\em IEEE Trans.
  Signal Process.}, vol.~64, no.~14, pp.~3775--3789, 2016.

\bibitem{ChenVSK:15}
S.~Chen, R.~Varma, A.~Sandryhaila, and J.~Kova{\v c}evi{\'c}, ``Discrete signal
  processing on graphs: {S}ampling theory,'' {\em IEEE Trans. Signal Process.},
  vol.~63, pp.~6510 -- 6523, Aug. 2015.

\bibitem{ChenVSK:15c}
S.~Chen, R.~Varma, A.~Singh, and J.~Kova{\v c}evi{\'c}, ``Signal recovery on
  graphs: Fundamental limits of sampling strategies,'' {\em IEEE Trans. Signal
  Inf. Process. Netw.}, vol.~2, no.~4, pp.~539--554, 2016.

\bibitem{ChenSMK:14}
S.~Chen, A.~Sandryhaila, J.~M.~F. Moura, and J.~Kova{\v c}evi{\'c}, ``Signal
  recovery on graphs: {Variation} minimization,'' {\em IEEE Trans. Signal
  Process.}, vol.~63, pp.~4609--4624, Sept. 2015.

\bibitem{ChenCRBGK:13}
S.~Chen, F.~Cerda, P.~Rizzo, J.~Bielak, J.~H. Garrett, and J.~Kova{\v
  c}evi{\'c}, ``Semi-supervised multiresolution classification using adaptive
  graph filtering with application to indirect bridge structural health
  monitoring,'' {\em IEEE Trans. Signal Process.}, vol.~62, pp.~2879--2893,
  June 2014.

\bibitem{NarangGO:13}
S.~K. Narang, A.~Gadde, and A.~Ortega, ``Signal processing techniques for
  interpolation in graph structured data,'' in {\em Proc. IEEE Int. Conf.
  Acoust., Speech, Signal Process.}, (Vancouver), pp.~5445--5449, May 2013.

\bibitem{ZhuM:12}
X.~Zhu and M.~Rabbat, ``Approximating signals supported on graphs,'' in {\em
  Proc. IEEE Int. Conf. Acoust., Speech, Signal Process.}, (Kyoto, Japan),
  pp.~3921 -- 3924, Mar. 2012.

\bibitem{ThanouSF:14}
D.~Thanou, D.~I. Shuman, and P.~Frossard, ``Learning parametric dictionaries
  for signals on graphs,'' {\em IEEE Trans. Signal Process.}, vol.~62,
  pp.~3849--3862, June 2014.

\bibitem{ChenVSK:15h}
S.~Chen, R.~Varma, A.~Singh, and J.~Kova{\v c}evi{\'c}, ``Signal
  representations on graphs: {T}ools and applications,'' {\em
  arXiv:1512.05406}, 2015.

\bibitem{AgaskarL:13}
A.~Agaskar and Y.~M. Lu, ``A spectral graph uncertainty principle,'' {\em IEEE
  Trans. Inf. Theory}, vol.~59, pp.~4338--4356, July 2013.

\bibitem{TsitsveroBL:15}
M.~Tsitsvero, S.~Barbarossa, and P.~D. Lorenzo, ``Signals on graphs:
  Uncertainty principle and sampling,'' {\em IEEE Trans. Signal Process.},
  vol.~64, no.~18, pp.~4845--4860, 2016.

\bibitem{NarangSO:10}
S.~K. Narang, G.~Shen, and A.~Ortega, ``Unidirectional graph-based wavelet
  transforms for efficient data gathering in sensor networks,'' in {\em Proc.
  IEEE Int. Conf. Acoust., Speech, Signal Process.}, (Dallas, TX),
  pp.~2902--2905, Mar. 2010.

\bibitem{HammondVG:11}
D.~K. Hammond, P.~Vandergheynst, and R.~Gribonval, ``Wavelets on graphs via
  spectral graph theory,'' {\em Appl. Comput. Harmon. Anal.}, vol.~30,
  pp.~129--150, Mar. 2011.

\bibitem{ShumanRV:15}
D.~I. Shuman, B.~Ricaud, and P.~Vandergheynst, ``Vertex-frequency analysis on
  graphs,'' {\em Appl. Comput. Harmon. Anal.}, vol.~40, no.~2, pp.~260--291,
  2016.

\bibitem{teke2016extending}
O.~Teke and P.~P. Vaidyanathan, ``Extending classical multirate signal
  processing theory to graphs¡ªpart {I}: Fundamentals,'' {\em IEEE Trans.
  Signal Process.}, vol.~65, no.~2, pp.~409--422, 2016.

\bibitem{teke2016extending2}
O.~Teke and P.~P. Vaidyanathan, ``Extending classical multirate signal
  processing theory to graphs¡ªpart {II}: M-channel filter banks,'' {\em IEEE
  Trans. Signal Process.}, vol.~65, no.~2, pp.~423--437, 2016.

\bibitem{oh2005tracking}
S.~Oh and S.~Sastry, ``Tracking on a graph,'' in {\em Proc. ACM/IEEE Int. Conf.
  Information Process. Sensor Netw.}, p.~26, IEEE Press, 2005.

\bibitem{Tremblay:14}
N.~Tremblay and P.~Borgnat, ``Graph wavelets for multiscale community mining,''
  {\em IEEE Trans. Signal Process.}, vol.~62, pp.~5227--5239, Oct. 2014.

\bibitem{DongFVN:14}
X.~Dong, P.~Frossard, P.~Vandergheynst, and N.~Nefedov, ``Clustering on
  multi-layer graphs via subspace analysis on {G}rassmann manifolds,'' {\em
  IEEE Trans. Signal Process.}, vol.~62, pp.~905--918, Feb. 2014.

\bibitem{ChenO:14}
P.-Y. Chen and A.~Hero, ``Local {F}iedler vector centrality for detection of
  deep and overlapping communities in networks,'' in {\em Proc. IEEE Int. Conf.
  Acoust., Speech, Signal Process.}, (Florence), pp.~1120--1124, 2014.

\bibitem{lim2014slashburn}
Y.~Lim, U.~Kang, and C.~Faloutsos, ``Slashburn: Graph compression and mining
  beyond caveman communities,'' {\em IEEE Trans. Knowl. Data Eng.}, vol.~26,
  no.~12, pp.~3077--3089, 2014.

\bibitem{lafon2006diffusion}
S.~Lafon and A.~B. Lee, ``Diffusion maps and coarse-graining: A unified
  framework for dimensionality reduction, graph partitioning, and data set
  parameterization,'' {\em IEEE Trans. Pattern Anal. Mach. Intell.}, vol.~28,
  no.~9, pp.~1393--1403, 2006.

\bibitem{shuman2016multiscale}
D.~I. Shuman, M.~J. Faraji, and P.~Vandergheynst, ``A multiscale pyramid
  transform for graph signals,'' {\em IEEE Trans. Signal Process.}, vol.~64,
  no.~8, pp.~2119--2134, 2016.

\bibitem{liu2014graph}
P.~Liu, X.~Wang, and Y.~Gu, ``Graph signal coarsening: Dimensionality reduction
  in irregular domain,'' in {\em Proc. GlobalSIP 2014}, pp.~798--802, IEEE,
  2014.

\bibitem{jung2016random}
J.~Jung, K.~Shin, L.~Sael, and U.~Kang, ``Random walk with restart on large
  graphs using block elimination,'' {\em ACM Trans. Database Syst.}, vol.~41,
  no.~2, p.~12, 2016.

\bibitem{leskovec2010kronecker}
J.~Leskovec, D.~Chakrabarti, J.~Kleinberg, C.~Faloutsos, and Z.~Ghahramani,
  ``Kronecker graphs: An approach to modeling networks,'' {\em J. Mach. Learn.
  Res.}, vol.~11, pp.~985--1042, 2010.

\bibitem{koutra2015summarize}
D.~Koutra, U.~Kang, J.~Vreeken, and C.~Faloutsos, ``Summarizing and
  understanding large graphs,'' {\em Statistical Analysis and Data Mining: The
  ASA Data Science Journal}, vol.~8, no.~3, pp.~183--202, 2015.

\bibitem{viterbi1967error}
A.~Viterbi, ``Error bounds for convolutional codes and an asymptotically
  optimum decoding algorithm,'' {\em IEEE Trans. Inf. Theory}, vol.~13, no.~2,
  pp.~260--269, 1967.

\bibitem{bettstetter2002minimum}
C.~Bettstetter, ``On the minimum node degree and connectivity of a wireless
  multihop network,'' in {\em Proceedings of the 3rd ACM international
  symposium on Mobile ad hoc networking \& computing}, pp.~80--91, ACM, 2002.

\bibitem{leskovec2005graphs}
J.~Leskovec, J.~Kleinberg, and C.~Faloutsos, ``Graphs over time: densification
  laws, shrinking diameters and possible explanations,'' in {\em Proc. ACM
  SIGKDD. Int. Conf. Knowl. Discovery Data Mining}, pp.~177--187, ACM, 2005.

\bibitem{CastroCD:11}
E.~Arias-Castro, E.~J. Cand{\`e}s, and A.~Durand, ``Detection of an anomalous
  cluster in a network,'' {\em The Annals of Statistics}, vol.~39, no.~1,
  p.~278¨C304, 2011.

\bibitem{HuCSFLL:13}
C.~Hu, L.~Cheng, J.~Sepulcre, G.~E. Fakhri, Y.~M. Lu, and Q.~Li, ``Matched
  signal detection on graphs: Theory and application to brain network
  classification,'' in {\em Proc. 23rd International Conference on Information
  Processing in Medical Imaging}, (Asilomara, CA), 2013.

\bibitem{SharpnackRS:13}
J.~Sharpnack, A.~Rinaldo, and A.~Singh, ``Changepoint detection over graphs
  with the spectral scan statistic,'' in {\em Artifical Intelligence and
  Statistics (AISTATS)}, 2013.

\bibitem{SharpnackKS:13}
J.~Sharpnack, A.~Krishnamurthy, and A.~Singh, ``Detecting activations over
  graphs using spanning tree wavelet bases,'' in {\em AISTATS}, (Scottsdale,
  AZ), Apr. 2013.

\bibitem{SharpnackKS:13a}
J.~Sharpnack, A.~Krishnamurthy, and A.~Singh, ``Near-optimal anomaly detection
  in graphs using lovasz extended scan statistic,'' in {\em Neural Information
  Processing Systems (NIPS)}, 2013.

\bibitem{ChenYZSK:16}
S.~Chen, Y.~Yang, S.~Zong, A.~Singh, and J.~Kova{\v c}evi{\'c}, ``Detecting
  structure-correlated attributes on graphs,'' {\em arXiv:1604.00657}, 2016.

\bibitem{agaskar2013detecting}
A.~Agaskar and Y.~M. Lu, ``Detecting random walks hidden in noise: Phase
  transition on large graphs,'' in {\em Proc. IEEE Int. Conf. Acoust., Speech,
  Signal Process.}, pp.~6377--6381, IEEE, 2013.

\bibitem{ting2006near}
M.~Ting, A.~O. Hero, D.~Rugar, C.-Y. Yip, and J.~A. Fessler, ``Near-optimal
  signal detection for finite-state markov signals with application to magnetic
  resonance force microscopy,'' {\em IEEE Trans. Signal Process.}, vol.~54,
  no.~6, pp.~2049--2062, 2006.

\bibitem{SinopoliSFPJS:04}
B.~Sinopoli, L.~Schenato, M.~Franceschetti, K.~Poolla, M.~I. Jordan, and S.~S.
  Sastry, ``Kalman filtering with intermittent observations,'' {\em IEEE Trans.
  Autom. Control}, vol.~49, pp.~1453--1464, Sept. 2004.

\bibitem{GordonN:93}
N.~Gordon, D.~Salmond, and A.~Smith, ``Novel approach to
  nonlinear/non-{G}aussian {B}ayesian state estimation,'' {\em IEEE Proc. Radar
  and Signal Process.}, vol.~140, no.~2, pp.~107--113, 1993.

\bibitem{lucas1981iterative}
B.~D. Lucas and T.~Kanade, ``An iterative image registration technique with an
  application to stereo vision,'' 1981.

\bibitem{hadjantonakis2004dynamic}
A.-K. Hadjantonakis and V.~E. Papaioannou, ``Dynamic in vivo imaging and cell
  tracking using a histone fluorescent protein fusion in mice,'' {\em BMC
  biotechnology}, vol.~4, no.~1, p.~33, 2004.

\bibitem{le2001diffusion}
D.~Le~Bihan, J.-F. Mangin, C.~Poupon, C.~A. Clark, S.~Pappata, N.~Molko, and
  H.~Chabriat, ``Diffusion tensor imaging: concepts and applications,'' {\em
  Journal of magnetic resonance imaging}, vol.~13, no.~4, pp.~534--546, 2001.

\bibitem{betzel2016multi}
R.~F. Betzel and D.~S. Bassett, ``Multi-scale brain networks,'' {\em
  NeuroImage}, 2016.

\bibitem{penrose2003random}
M.~Penrose, {\em Random geometric graphs}.
\newblock No.~5, Oxford University Press, 2003.

\bibitem{fortunato2010community}
S.~Fortunato, ``Community detection in graphs,'' {\em Physics reports},
  vol.~486, no.~3, pp.~75--174, 2010.

\bibitem{zeitouni1992generalized}
O.~Zeitouni, J.~Ziv, and N.~Merhav, ``When is the generalized likelihood ratio
  test optimal?,'' {\em IEEE Transactions on Information Theory}, vol.~38,
  no.~5, pp.~1597--1602, 1992.

\bibitem{du2016convergence}
J.~Du, S.~Ma, Y.-C. Wu, S.~Kar, and J.~M.~F. Moura, ``Convergence analysis of
  distributed inference with vector-valued gaussian belief propagation,'' {\em
  arXiv:1611.02010}, 2016.

\bibitem{du2017convergence}
J.~Du, S.~Ma, Y.-C. Wu, S.~Kar, and J.~M.~F. Moura, ``Convergence analysis of
  the information matrix in gaussian belief propagation,'' in {\em IEEE
  International Conference on Acoustics, Speech and Signal Processing
  (ICASSP)}, 2017.

\bibitem{toumpis2004large}
S.~Toumpis and A.~J. Goldsmith, ``Large wireless networks under fading,
  mobility, and delay constraints,'' in {\em INFOCOM 2004}, vol.~1, IEEE, 2004.

\bibitem{karypis1998fast}
G.~Karypis and V.~Kumar, ``A fast and high quality multilevel scheme for
  partitioning irregular graphs,'' {\em SIAM Journal on scientific Computing},
  vol.~20, no.~1, pp.~359--392, 1998.

\bibitem{blondel2008fast}
V.~D. Blondel, J.-L. Guillaume, R.~Lambiotte, and E.~Lefebvre, ``Fast unfolding
  of communities in large networks,'' {\em Journal of statistical mechanics:
  theory and experiment}, vol.~2008, no.~10, p.~P10008, 2008.

\bibitem{von2007tutorial}
U.~Von~Luxburg, ``A tutorial on spectral clustering,'' {\em Statistics and
  computing}, vol.~17, no.~4, pp.~395--416, 2007.

\end{thebibliography}

\section{Supplementary: Proof of Lemma~\ref{lmm:pro_upb}}\label{app:pro_upb}
Define $W_t=\u_t(\hat{\V}_t)$ and $U_t=\u_t(\V^*_t)$. The probability of this event can be upper-bounded by
\begin{equation}\small
  \Pr(S(\hat{\Pd})\ge S(\Pd^*))=\Pr\l(\sum_{t=1}^TW_t\ge \sum_{t=1}^TU_t\r),
\end{equation}
where note that at some positions, the two paths may overlap. Using the Markov inequality, for all $s>0$,
\begin{equation}\label{eqn:sc_ge_st}\small
\begin{split}
  &\Pr\l(S(\hat{\Pd})\ge S(\Pd^*)\r)=\Pr\l(\exp\l(s\sum_{t=1}^T(W_t-U_t)\r)\ge 1\r)\\
  &\le \min_{s>0}\Ep\l[\exp\l(s\sum_{t=1}^T(W_t-U_t)\r) \r]\overset{(a)}{=}\min_{s>0}\prod_{t\in \Delta}\Ep\l[e^{sW_t}\r]\Ep\l[e^{-sU_t}\r],
\end{split}
\end{equation}
where $\Delta\subset\{1,2, \ldots  T\}$ in equality (a) denotes the set of time points when the two paths do not overlap, i.e., $\hat{\V}_t\neq\V^*_t$. From the definitions of $W_t$, we know $W_t$ is the maximum of $|\hat{\V}_t|$ i.i.d. random variables $W_i^\text{off}$, where each $W_i^\text{off}\overset{\D}{=}W^\text{off}\sim\N(0,\sigma^2)$. We use the following lemma to upper-bound the moment-generating function $\Ep\l[e^{sW_t}\r]$ of $W_t$.
\begin{lemma}\label{lmm:qfunc}
Suppose $X_1,X_2,\ldots X_k$ are i.i.d. Gaussian random variables with mean zero and variance $\sigma^2$. Denote by $X_\textrm{max}=\max\{X_1,X_2,\ldots X_k\}$. Then\footnote{\textcolor{black}{An astute reader may be confused by the $\min_{l\ge k}$ in \eqref{eqn:min} because $e^{sX_\textrm{max}}$ is monotonically increasing in $l$. However, the right hand side of \eqref{eqn:min} is not necessarily monotone in $l$. Nonetheless, the $\min_{l\ge k}$ is only a proof trick to make the function $\psi(s,k)$ in \eqref{eqn:fsl} monotonically increasing in $k$. Finally we will assign $l=k$ to obtain the upper desired bound.}},
\begin{equation}\label{eqn:min}\small
\begin{split}
\Ep&\l[e^{sX_\textrm{max}}\r]\le \min_{l\ge k}\min_{\eta\in[0,1]}l\eta^{l-1}e^{\frac{1}{2}s^2\sigma^2} Q(Q^{-1}(\eta/\sigma)-s\sigma^2)\\
&+\sqrt{ \frac{l^2}{2l-1} (1-\eta^{2l-1})}e^{s^2\sigma^2}\sqrt{ Q(Q^{-1}(\eta/\sigma)-2s\sigma^2)}.
\end{split}
\end{equation}
\end{lemma}

\begin{IEEEproof}
Denote by $F(x)$ the c.d.f. (cumulative distribution function) of $X_i,i=1,2,\ldots k$. Denote by $\phi(x)$ the p.d.f. (probabilistic distribution function) of $X_i,i=1,2,\ldots k$. Then, the p.d.f. of $X_\textrm{max}$ is $kF(x)^{k-1}\phi(x)$. Note that the maximum of $l\ge k$ random variables that have the same distribution $\phi(x)$ as $X_i$ have a larger moment generating function than $\Ep\l[e^{sX_\textrm{max}}\r]$. Therefore, for any $\eta\in [0,1]$ and $l\ge k$,
\begin{equation}\label{eqn:qfunc0}\small
\begin{split}
&\Ep\l[e^{sX_\textrm{max}}\r]\le\int_{-\infty}^{\infty} lF(x)^{l-1}\phi(x)e^{sx} dx\\
= & \int_{-\infty}^{F(x)= \eta} lF(x)^{l-1}\phi(x)e^{sx} dx
     +\int_{F(x)= \eta}^{\infty} lF(x)^{l-1}\phi(x)e^{sx} dx\\
\overset{(a)}{\le} & \int_{-\infty}^{F(x)= \eta} l\eta^{l-1}\phi(x)e^{sx} dx\\
    & +\sqrt{\int_{F(x)= \eta}^{\infty} l^2F(x)^{2l-2}\phi(x) dx}\cdot\sqrt{\int_{F(x)= \eta}^{\infty} \phi(x)e^{2sx}dx}\\
= & l\eta^{l-1}\int_{-\infty}^{F(x)= \eta} \phi(x)e^{sx} dx\\
 & +\sqrt{\int_{F(x)= \eta}^{\infty} l^2 dF(x)^{2l-1}}\sqrt{\int_{F(x)= \eta}^{\infty} \phi(x)e^{2sx}dx}\\
= & l\eta^{l-1}\int_{-\infty}^{F(x)= \eta} \phi(x)e^{sx} dx\\
& +\sqrt{ \frac{l^2}{2l-1} (1-\eta^{2l-1})}\sqrt{\int_{F(x)= \eta}^{\infty} \phi(x)e^{2sx}dx},
\end{split}
\end{equation}
where (a) follows from the Cauchy-Schwartz inequality. Notice that
\begin{equation}\label{eqn:qfunc1}\small
\begin{split}
&\int_{-\infty}^{F(x)= \eta} \phi(x)e^{sx} dx\\
=&\frac{1}{\sqrt{2\pi\sigma^2}}\int_{-\infty}^{x=F^{-1}(\eta)}e^{-\frac{1}{2\sigma^2}x^2}e^{sx}dx\\
=&\frac{1}{\sqrt{2\pi\sigma^2}}\int_{-\infty}^{x=F^{-1}(\eta)}e^{-\frac{1}{2\sigma^2}(x-s\sigma^2)^2}e^{\frac{1}{2}s^2\sigma^2}dx\\
=&e^{\frac{1}{2}s^2\sigma^2} \frac{1}{\sqrt{2\pi\sigma^2}}\int_{-\infty}^{x=F^{-1}(\eta)-s\sigma^2}e^{-\frac{1}{2\sigma^2}x^2}dx\\
=&e^{\frac{1}{2}s^2\sigma^2} Q(F^{-1}(\eta)-s\sigma^2)=e^{\frac{1}{2}s^2\sigma^2} Q(Q^{-1}(\eta/\sigma)-s\sigma^2),
\end{split}
\end{equation}
where $Q(\cdot)$ is the Q-function $ Q(x) = \frac{1}{\sqrt{2\pi}} \int_x^\infty \exp\left(-\frac{u^2}{2}\right) \, du$. Also notice that
\begin{equation}\label{eqn:qfunc2}\small
\begin{split}
&\int_{F(x)= \eta}^{\infty} \phi(x)e^{2sx}dx
=\frac{1}{\sqrt{2\pi\sigma^2}}\int_{x=F^{-1}(\eta)}^{\infty}e^{-\frac{1}{2\sigma^2}x^2}e^{2sx}dx\\
=&\frac{1}{\sqrt{2\pi\sigma^2}}\int_{x=F^{-1}(\eta)}^{\infty}e^{-\frac{1}{2\sigma^2}(x-2s\sigma^2)^2}e^{2s^2\sigma^2}dx\\
=&e^{2s^2\sigma^2} \frac{1}{\sqrt{2\pi\sigma^2}}\int_{x=F^{-1}(\eta)-2s\sigma^2}^{\infty}e^{-\frac{1}{2\sigma^2}x^2}dx\\
=&e^{2s^2\sigma^2} Q(F^{-1}(\eta)-2s\sigma^2)=e^{2s^2\sigma^2} Q(Q^{-1}(\eta/\sigma)-2s\sigma^2).
\end{split}
\end{equation}
Plugging in \eqref{eqn:qfunc1} and \eqref{eqn:qfunc2} into \eqref{eqn:qfunc0}, we have
\begin{equation}\label{eqn:qfunc3}\small
\begin{split}
&\Ep\l[e^{sX_\textrm{max}}\r]
\le l\eta^{l-1}e^{\frac{1}{2}s^2\sigma^2} Q(Q^{-1}(\eta/\sigma)-s\sigma^2)\\
&+\sqrt{ \frac{l^2}{2l-1} (1-\eta^{2l-1})}e^{s^2\sigma^2}\sqrt{ Q(Q^{-1}(\eta/\sigma)-2s\sigma^2)}.
\end{split}
\end{equation}
\end{IEEEproof}

Using Lemma \ref{lmm:qfunc}, we have
\begin{equation}\label{eqn:ineq_off}\small
\begin{split}
  &\Ep\l[e^{sW_t}\r]\le \min_{l\ge |\hat{\V}_t|}\min_{\eta\in[0,1]}l\eta^{l-1}e^{\frac{1}{2}s^2\sigma^2} Q(Q^{-1}(\eta/\sigma)-s\sigma^2)\\
  &\quad+\sqrt{ \frac{l^2}{2l-1} (1-\eta^{2l-1})}e^{s^2\sigma^2}\sqrt{ Q(Q^{-1}(\eta/\sigma)-2s\sigma^2)},
\end{split}
\end{equation}
where $l=|\hat{\V}_t|$.

From the definition of $U_t$, $U_t$ is always greater than a random variable that is distributed the same as $U^\text{on}\sim\N(\mu,\sigma^2)$. Therefore,
\begin{equation}\label{eqn:ineq_on}\small
  \Ep\l[e^{-sU_t}\r]\le \Ep\l[e^{-sU^\text{on}}\r]=e^{\frac{1}{2}s^2\sigma^2-\mu s}.
\end{equation}
Therefore, from \eqref{eqn:sc_ge_st},~\eqref{eqn:ineq_off} and \eqref{eqn:ineq_on},
\begin{equation}\small
\begin{split}
\Pr\l(S(\hat{\Pd})\ge S(\Pd^*)\r)\le \min_{s>0} \prod_{t\in \Delta} \psi(s,|\hat{\V}_t|)\le \prod_{t\in \Delta} \psi(s^*,|\hat{\V}_t|),
\end{split}
\end{equation}
where $s^*$ is chosen to be $s^*=\frac{\mu}{2\sigma^2}$ and
\begin{equation}\label{eqn:fsl}\small
\begin{split}
&\psi(s,k):=\min_{l\ge k}\min_{\eta\in[0,1]}l\eta^{l-1}e^{s^2\sigma^2-\mu s} Q(Q^{-1}(\eta/\sigma)-s\sigma^2)\\
&+\sqrt{ \frac{l^2}{2l-1} (1-\eta^{2l-1})}e^{\frac{3}{2}s^2\sigma^2-\mu s}\sqrt{ Q(Q^{-1}(\eta/\sigma)-2s\sigma^2)}.
\end{split}
\end{equation}
Therefore, we complete the proof by upper bounding $\psi(s^*,|\hat{\V}_t|)$ with $\theta(s^*,|\hat{\V}_t|)$ defined in \eqref{eqn:funcf}.

\section{\textcolor{black}{Supplementary: An Extension to Localizing Multiple Path-Signals}}\label{sec:multi_path}

Consider the problem of localizing multiple (possibly overlapping) paths from noisy observations. Suppose there are $k>1$ deterministic but unknown connected paths $\p_j=(v_1^j,v_2^j,\ldots v_T^j),j=1,2\ldots k$. The multi-path-signal $\x_t$ is defined as follows: $\x_t(v)=\mu$ for all $v\in \{v_t^1,v_t^2,\ldots v_t^k\}$, i.e., if $v$ is on at least one path (two paths may overlap at $v$) at time $t$, otherwise $\x_t(v)=0$. The observation model is still defined as
\begin{equation}
\y_t=\x_t+\w_t,t=1,2, \ldots T,
\end{equation}
where the noise $\w_t\sim \N( \0, \sigma^2 \Id)$. Our goal is to localize the set of nodes $\{v_t^1,v_t^2,\ldots v_t^k\}$ for each time point $t$. We do not require to recover the index of the path that each node belongs to.

The generalization of the multiscale Viterbi algorithm to multiple paths is shown in Algorithm \ref{alg:dp_multiple}. The main intuition behind Algorithm~\ref{alg:dp_multiple} is that the $k$ paths can be found sequentially. After one path is found, the activated path-signal is subtracted from the signal observations $\y_t$ and the search for the next path begins. \textcolor{black}{The empirical performance of the proposed multi-path multiscale Viterbi algorithm degrades as the number of paths increases, because when many paths overlap with each other, subtracting a path-signal from an overlapping path-signal makes the latter one disconnected. Therefore, when localizing multiple path-signals in a sequential way, the path localization error accumulates. A thorough study on the performance of this extension is meaningful}.

\begin{algorithm}\caption{Multiscale Viterbi Decoding for Multi-Path-Signal Localization}\label{alg:dp_multiple}
\textbf{INPUT}: A graph $\G=(\V,\E)$ and graph signal observations $\y_t,t=1,2, \ldots T$.\\
\textbf{OUTPUT}: $T$ nodes sets $\{v_t^1,v_t^2,\ldots v_t^k\},t=1,2,\ldots T$.\\
\textbf{INITIALIZE}: Set $\w_t=\y_t$.\\
\textbf{FOR} $j$ from 1 to $k$
\begin{itemize}
\item Call Algorithm~\ref{alg:dp_multi_scale} with inputs $\G=(\V,\E)$ and $\w_t,t=1,2, \ldots T$ to obtain a path estimate $\hat{\p}_j=(\hat{v}_1^j,\hat{v}_2^j,\ldots \hat{v}_T^j)$.
\item Set $\w_t(\hat{v}_t^j)\leftarrow \w_t(\hat{v}_t^j)-\mu$.
\end{itemize}
\textbf{OUTPUT}: The $T$ nodes sets $\{v_t^1,v_t^2,\ldots v_t^k\}, t=1,2,\ldots T$.
\end{algorithm}

\begin{figure}
  \centering
  \includegraphics[scale=0.4]{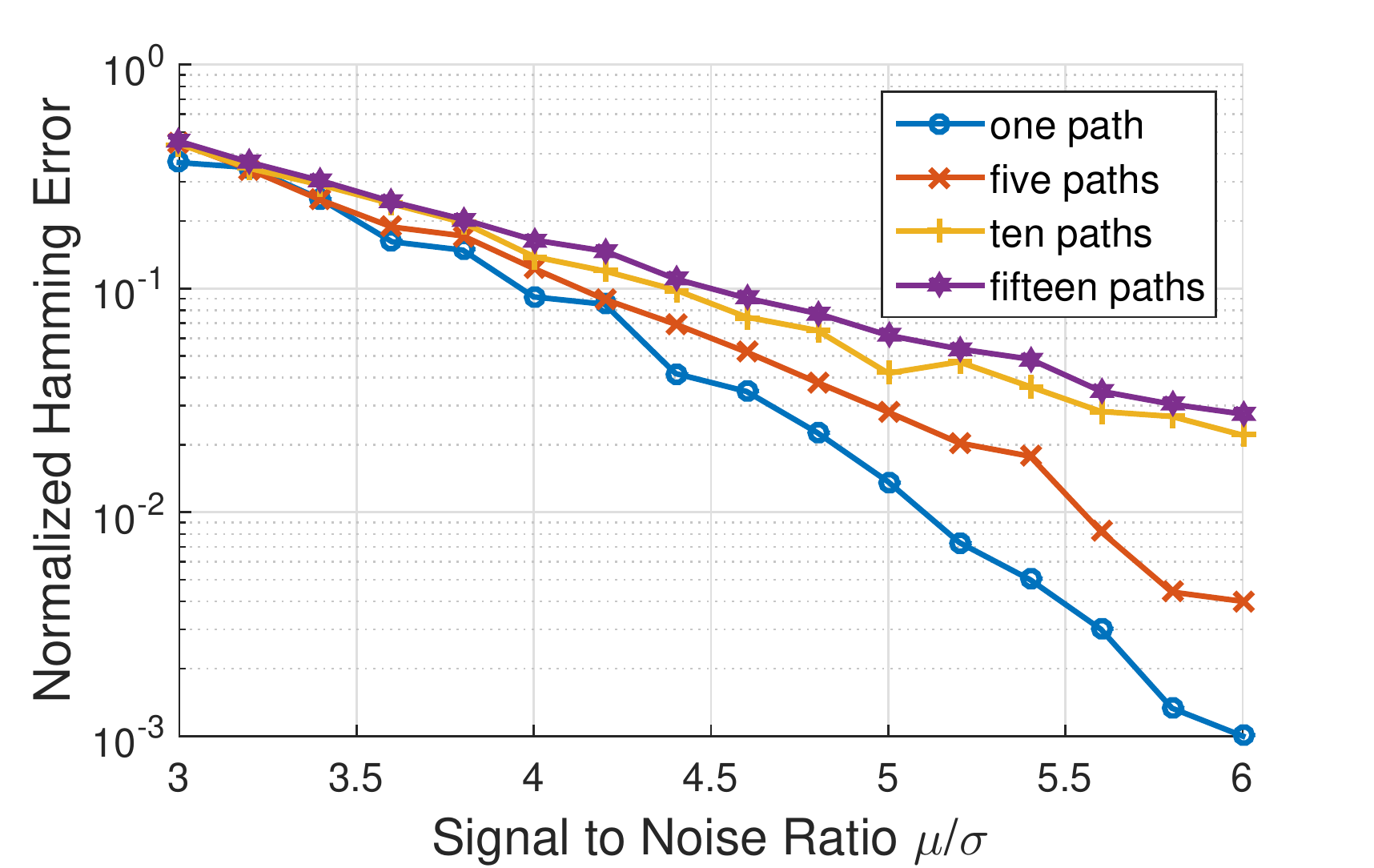}\\
  \caption{Simulation results on the Hamming distance between the path estimates and the true paths for different number of paths.}\label{fig:geo_multipath}
\end{figure}

\textcolor{black}{Finally, we test the multi-path Viterbi decoding algorithm on the same random geometric graph (Algorithm \ref{alg:dp_multiple}) with 900 clusters. The results are shown in Fig. \ref{fig:geo_multipath}. Each curve represents the average normalized Hamming distance for a particular number of paths. Note that the localization result is not good enough when we try to localize ten paths at the same time. This is because when we try to localize ten paths, we have to sequentially carry out the multiscale Viterbi decoding algorithm (Algorithm~\ref{alg:dp_multi_scale}) on the super-graph formed by 900 clusters. This means we effectively search for 10 paths in a graph with 900 nodes, which may have many overlappings or crossings.}

\section{Supplementary: Proof of Corollary~\ref{cor:dis_Hamming}}\label{app:Hamming}
From Theorem~\ref{thm:Hamming}, we can replace $|\hat{\V}_t|$ with $s_m$ in~\eqref{eqn:dis_Hamming}. In this proof, we will only count the paths $\hat{\Pd}=(\hat{\V}_1,\hat{\V}_2, \ldots \hat{\V}_T)$ s.t. $D_H(\Pd^*,\hat{\Pd})\ge \delta T$ for some constant $\delta$. We denote this set of paths by $\S_\delta$. Then, from Theorem~\ref{thm:Hamming},
\begin{equation}\small
\Ep\l[D_H(\Pd^*,\hat{\Pd})\r]\le \delta T+\sum_{\hat{\Pd}\in\S_\delta}\l[D_H(\Pd^*,\hat{\Pd})\prod_{t\in \Delta(\hat{\Pd})}\theta\l(\frac{\mu}{2\sigma^2},|\hat{\V}_t|\r)\r],
\end{equation}
where $D_H(\Pd^*,\hat{\Pd})=|\Delta(\hat{\Pd})|$. From the definition of $\theta(\cdot,\cdot)$ in \eqref{eqn:funcf}, we have
\begin{equation}\small
\begin{split}
&\theta\l(\frac{\mu}{2\sigma^2},|\hat{\V}_t|\r)
\overset{(a)}{=}\min_{\eta\in[0,1]}l\eta^{l-1}e^{s^2\sigma^2-\mu s} Q(Q^{-1}(\eta/\sigma)-s\sigma^2)\\
&+\sqrt{ \frac{l^2}{2l-1} (1-\eta^{2l-1})}e^{\frac{3}{2}s^2\sigma^2-\mu s}\sqrt{ Q(Q^{-1}(\eta/\sigma)-2s\sigma^2)}\\
\overset{(b)}{\le}& le^{s^2\sigma^2-\mu s}Q(Q^{-1}(1/\sigma)-s\sigma^2)\\
\le&le^{s^2\sigma^2-\mu s}\\
\overset{(c)}{\le}& s_m\exp\l(-\frac{\mu^2}{4\sigma^2}\r),
\end{split}
\end{equation}
where $l=|\hat{\V}_t|$ and $s=\frac{\mu}{2\sigma^2}$ in (a), (b) is obtained by setting $\eta=1$, and (c) is obtained by plugging in $s=\frac{\mu}{2\sigma^2}$ and $l=|\hat{\V}_t|\le s_m$. Thus, we have
\begin{equation}\small
\Ep\l[D_H(\Pd^*,\hat{\Pd})\r]\le \delta T+\sum_{\hat{\Pd}\in\S_\delta}\l[D_H(\Pd^*,\hat{\Pd})\prod_{t\in \Delta(\hat{\Pd})}\exp\l(-\frac{\mu^2}{4\sigma^2}\r)s_m\r].
\end{equation}
Note that $\Delta(\hat{\Pd})$ is defined as the subset of time when $\hat{\V}_t\neq\V^*_t$. Therefore, if we define
\begin{equation}\small
  c:=\exp\l(-\frac{\mu^2}{4\sigma^2}\r)s_m,
\end{equation}
we have
\[\small\begin{split}
&\Ep\l[D_H(\Pd^*,\hat{\Pd})\r]\le \delta T+\sum_{\hat{\Pd}\in\S_\delta}\l[D_H(\Pd^*,\hat{\Pd})\prod_{t\in \Delta(\hat{\Pd})}c\r]\\
=&\delta T+\sum_{\hat{\Pd}\in\S_\delta}\l[D_H(\Pd^*,\hat{\Pd})c^{D_H(\Pd^*,\hat{\Pd})}\r]
=\delta T+\sum_{D=\delta T}^T \N_D\cdot Dc^D,
\end{split}\]
where $\N_D$ is the number of paths $\hat{\Pd}$ such that $D_H(\hat{\Pd},\Pd^*)=D$. The total number of paths that has distance $D$ away from the true path is upper-bounded by
\begin{equation}\small
  \N_D<\binom{T}{D} 9^D,
\end{equation}
where the first term $\binom{T}{D}$ denotes the possible positions of the $D$ time points $t$ such that $\hat{\V}_t\neq\V^*_t$ in $T$ time points, and the second term $9^D$ is the upper bound on the number of paths that differs from the true path on particular $D$ positions. Therefore
\[\small\begin{split}
\Ep\l[D_H(\Pd^*,\hat{\Pd})\r]\le&\delta T+\sum_{D=\delta T}^T \binom{T}{D} 9^D\cdot Dc^D\\
\le &\delta T+T\sum_{D=\delta T}^T \binom{T}{D} (9c)^D.\\
  =&\delta T+ \left[\sum_{D=\delta T} ^T \binom{T}{T-D} \left(\frac{1}{9c}\right)^{T-D}\right]\cdot(9c)^T\\
  =&\delta T+ \left[\sum_{U=0} ^{(1-\delta)T} \binom{T}{U} \left(\frac{1}{9c}\right)^{U}\right]\cdot(9c)^T\\
  \le&\delta T+ \left(\frac{1}{9c}\right)^{h_q(1-\delta)T}(9c)^T,
\end{split}\]
where the last inequality follows from an upper bound on the volume of a Hamming ball with radius $(1-\delta)T$ that holds when
\begin{equation}\small
  \delta\ge \frac{1}{q}:=9c,
\end{equation}
and the entropy function $h_q(p)=p\log_q(q-1)-p\log_q p-(1-p)\log_q(1-p)$. Therefore,
\begin{equation}\small
\begin{split}
  \Ep\l[D_H(\Pd^*,\hat{\Pd})\r]\le&\delta T+(9c)^{[1-h_q(1-\delta)]T},
\end{split}
\end{equation}
where $\delta$ has to satisfy $\delta\ge 9c$, where $c$ should be chosen such that $9c<1$. In other words, when $\delta\ge 9c\in[0,1]$, the term $(9c)^{[1-h_q(1-\delta)]T}$ goes to zero when $T\to \infty$. Thus,
\begin{equation}\small
  \lim_{T\to\infty}\Ep\l[D_H(\Pd^*,\hat{\Pd})\r]\le 9c T=9\exp\l(-\frac{\mu^2}{4\sigma^2}\r)s_m T,
\end{equation}
when $\exp\l(-\frac{\mu^2}{4\sigma^2}\r)s_m<\frac{1}{9}$, or $\mu/\sigma>2\sqrt{\log(9s_m)}$.

\end{document}